\documentclass[a4paper,11pt]{article}
\pdfoutput=1 

\usepackage{jcappub} 

\usepackage[T1]{fontenc} 

\title{\boldmath Probing the ionizing photon output of galaxies near cosmic dawn with the patchy kSZ effect}

\author[a]{Garett Lopez,}
\author[a]{Anson D'Aloisio,}
\author[b]{and Christopher Cain}

\affiliation[a]{Department of Physics and Astronomy, University of California, Riverside,
\\900 University Ave., Riverside, CA, 92521, USA}
\affiliation[b]{School of Earth and Space exploration, Arizona State University
\\Tempe, AZ 85281, USA}

\emailAdd{garett.lopez@ucr.edu}
\emailAdd{ansond@ucr.edu}
\emailAdd{clcain3@asu.edu}

\abstract{A key result from JWST’s first cycles is that galaxy formation was well underway by $z=10$. The implications of these early galaxies for reionization are less clear, however. The CMB is one of the few windows into the ionization state of the IGM during reionization’s first half, providing an important probe of the ionizing photon sources at those times.  Meanwhile, measurements of the Lyman-$\alpha$ forest in the spectra of high-$z$ quasars have improved to the level of tightly constraining the timing of reionization's end. In this paper, we use radiative transfer simulations to explore how measurements of the patchy kinetic Sunyaev Zel’dovich (pkSZ) effect, when combined with Lyman-$\alpha$ forest measurements, can be used to constrain the early stages of reionization and the nature of its sources.  For a given source model, we find that the amplitude of the pkSZ power spectra strongly correlates with the start time of reionization, and constrains the number of ionizing photons produced by the high-$z$ source population.  Allowing for variations in the source model, this correlation is weakened by a degeneracy between the reionization history and the effects of source clustering. However, we demonstrate two potential ways of breaking this degeneracy using: (1) measurements of large-scale fluctuations in the Ly$\alpha$ forest opacity at $z=5-6$, and/or; (2) the shape of the pkSZ power spectrum measured in future CMB surveys.  Models with highly clustered sources yield steeper slopes in the pkSZ power around $\ell = 3,000$, so measurements at additional angular scales can be used to break the history-clustering degeneracy.  Our results highlight how future pkSZ measurements will complement JWST observations to improve our understanding of the ionizing sources near cosmic dawn.}


\begin{document}

\maketitle
\flushbottom

\section{Introduction}
\label{sec:intro}

Over the past decade, quasar absorption spectra measurements have sharpened our understanding of reionization's last phases.   A number of lines of evidence point to the last $\lesssim 20\%$ of reionization occurring at $z<6$.  These include the observed evolution in the Lyman-$\alpha$ (Ly$\alpha$) forest mean flux and large-scale opacity fluctuations \cite{2019MNRAS.485L..24K, 2020MNRAS.491.1736K, 2020MNRAS.494.3080N}, dark gap statistics of the co-eval Ly$\alpha$ and Ly$\beta$ forests \cite{2021ApJ...923..223Z, 2022ApJ...932...76Z}, and rapid evolution in the mean free path of ionizing photons between $z=5-6$ \cite{2021MNRAS.508.1853B, 2023ApJ...955..115Z, 2023MNRAS.525.4093G, 2024ApJ...965..134D, 2024MNRAS.530.5209R, 2024MNRAS.533..676S}.  Most recently, Refs. \cite{2024arXiv240508885B, 2024arXiv240512275Z, 2024arXiv240512273S} reported evidence of damping wings at the edges of dark gaps in the co-eval Ly$\alpha$ and Ly$\beta$ forests at $z<6$, so far the most direct indication of ongoing reionization at those times.   In combination, these measurements have moved beyond providing just a lower bound in redshift for the end of reionization; they constrain the timing of it \cite{Bosman2021, 2023MNRAS.525.4093G, 2025PASA...42...49Q, 2025arXiv250515899C}.  The central focus of the current paper is how this constraint alters the interpretation of existing and future measurements of the kinetic Sunyaev-Zel'dovich (kSZ) signal from reionization.   

Comparatively little is known about reionization's earliest phases.  With the wavelength coverage and spectroscopy provided by the Near Infrared Camera and Spectrometer (NIRCam and NIRSpec), the James Webb Space Telescope (JWST) promises groundbreaking insights into the earliest sources of reionization.  JWST observations have already provided measurements of the cosmic rest-frame ultraviolet (UV) luminosity density ($\rho_{\rm UV}$) at redshifts up to $z\sim 14$ \cite{2023MNRAS.518.6011D, 2023MNRAS.523.1009B, 2023ApJ...948L..14C, 2023ApJ...951L...1P, 2023ApJ...954L..46L, 2024MNRAS.527.5004M, Finkelstein2024, Adams2024, 2025arXiv250100984W}.   These early studies have not yet converged on a coherent picture, likely owing to sample variance and/or low redshift interlopers \cite[e.g.][]{Adams2024, 2023MNRAS.523.1009B, 2023Natur.622..707A}.  In aggregate, they make clear that galaxies were forming just a few hundred million years after the Big Bang. The implications of these observations for reionization, however, are less clear.  The ionizing emissivity, or number of ionizing photons per unit comoving volume emitted into the intergalactic medium (IGM) at redshift $z$, is often modeled with the ansatz

\begin{equation}
\dot{N}_{\rm ion} = \langle f_{\rm esc} \xi_{\rm ion} \rangle \cdot \rho_{\rm UV},
\label{eq1}
\end{equation}
where $f_{\rm esc}$ is the escape fraction of ionizing photons, and $\xi_{\rm ion}$ is the ionizing efficiency, the conversion factor between a galaxy's non-ionizing UV luminosity and its rate of hydrogen-ionizing photon production.\footnote{The non-ionizing UV luminosity of a galaxy is typically measured at rest-frame 1500 \AA.} The brackets denote a UV-luminosity-weighted average over the galaxy population.  With JWST, ionizing efficiencies have been measured for samples of up to a few dozen galaxies at $z>5$ \cite{Endsley2023, Pahl2024, 2024MNRAS.527.6139S}. Again, the conclusions of these early studies are varied, with some reporting a factor of a few enhancement in $\xi_{\rm ion}$ over the value of $\log_{10}(\xi_{\rm ion}/\mathrm{erg~Hz}^{-1}) = 25.2$ commonly adopted before JWST (e.g. \cite{2015ApJ...802L..19R}).    The variation among these early results could be a result of redshift evolution and/or a correlation between $\xi_{\rm ion}$ and UV luminosity \cite{Endsley2023, 2024MNRAS.527.6139S, Pahl2024}. 

But even if we had a precise empirical model for $\xi_{\rm ion}$ and its main dependencies, the implications for reionization would still be difficult to glean for two reasons: (1) it is impossible to measure $f_{\rm esc}$ directly at these redshifts owing to the prohibitively large opacity of the IGM to ionizing photons; (2) much of the ionizing photon budget could have been sourced by galaxies too faint to be detected by JWST \cite[e.g.][]{Finklestein2019, 2024Natur.626..975A, 2025arXiv250100984W}.  Indeed, Ref. \cite{2024MNRAS.535L..37M} attempted to hash out the implications of the recent JWST results for reionization. They argued that even the {\it observed} high-$z$ galaxies with $M_{\rm UV} < -15$ should have been prolific producers and leakers of ionizing photons; so much so that reionization would have completed well before $z = 6$, in violation of Ly$\alpha$ forest constraints.  Their argument, however, relies on a model for $f_{\rm esc}$ that is based on a correlation between $f_{\rm esc}$ and UV-continuum slope measured at $z\sim 0$ \cite{2022MNRAS.517.5104C}, and a measurement of $\xi_{\rm ion}$ reported by Ref. \cite{2024MNRAS.527.6139S}, which is likely biased high by selection effects \cite{2024MNRAS.535.2998S}.     This discussion serves to highlight the large uncertainties still at play when attempting to infer the ionizing output of high-$z$ galaxies from their non-ionizing spectral properties (see also the detailed discussion of these nuances in Ref. \cite{2024arXiv240902989C}). 

Constraints on the ionization state of the IGM are a critical indirect probe of the ionizing emissivity of the galaxy population. Unfortunately, at $z> 8$, such constraints are still limited. To date, 21 cm experiments targeting the epoch of reionization have placed only upper limits on the $z \geq 8$ power spectrum~\citep{HERA2023,Nunhokee2025}. Observations of Ly$\alpha$ emission in high-$z$ galaxy spectra have begun to place constraints on reionization's early stages~\citep[e.g.][]{Umeda2023,Witstok2024,Mason2025}, but these are limited by poor statistics and model dependence~\citep{Hassan2021}.  The cosmic microwave background (CMB) provides integral constraints that stem from measurements of the EE power spectrum at low $\ell$ and from the contribution of reionization to the kSZ effect, the so-called patchy kSZ contribution (pkSZ)\footnote{In some works this is also termed the ``reionization kSZ'' or rkSZ. }. The former is typically expressed as the summary statistic $\tau_{\rm es}$, the electron scattering optical depth, although additional information is contained in the detailed shape of the EE power.  Measurements of $\tau_{\rm es}$ have almost always been interpreted as a constraint on reionization's midpoint.   Ref. \cite{2025arXiv250515899C} recently pointed out that when $\tau_{\rm es}$ is combined with Ly$\alpha$ forest constraints on the end of reionization, it becomes a measurement of reionization's {\it duration}.  Perhaps this could be exploited to glean insights into reionization's early phases. However, given the low value of $\tau_{\rm es} =  0.058 \pm 0.0062$ measured from Planck data \cite{Planck18, 2024A&A...682A..37T}, the results of Ref. \cite{2021MNRAS.508.2784W} suggest that even a cosmic-variance limited measurement of the low-$\ell$ EE power would be unlikely to discriminate between realistic models. 

The pkSZ signal provides an alternative path forward.  Ref. \cite{Reichardt2020} reported a $\ge 3\,\sigma$ detection of the total kSZ power at $\ell = 3,000$,
of $D_{3000} = 3.0\pm1.0 \mu$K$^2$  from the South Pole Telescope (SPT).  They measured the pkSZ component to be $D_{3000}^{\rm pkSZ} = 1.1^{+1.0}_{-0.7}\mu$K$^2$ using estimates of the post-reionization contribution from \cite{Calabrese2014, Shaw2012}.   During the preparation of this manuscript, Ref. \cite{Beringue2025-ka} measured a total kSZ power of $D_{3000} = 1.48_{-1.10}^{+0.71} \mu$K$^2$ with the Atacama Cosmology Telescope (ACT).\footnote{Ref. \cite{Beringue2025-ka} did not repot a pkSZ measurement. For reference, adopting a conservative (low) estimate of 0.85 $\mu$K$^2$ for the post-reionization contribution to the kSZ power from Ref. \cite{Shaw2012} yields a $2 \sigma$ upper limit on the pkSZ power of $D_{3000}^{\rm pkSZ}\lesssim$ 2.05 $\mu$K$^2$.} Forthcoming experiments such as Simon's Observatory and CMB-S4 Wide are forecasted to measure $D_{3000}^{\rm pkSZ}$ to a precision of $\sim \pm 0.2 \mu$K$^2$ (1-$\sigma$ limits), improving to $\sim \pm 0.1 \mu$K$^2$ when combined with future LiteBIRD constraints on $\tau_{\rm es}$ \cite{Jain2023-tk}.  In this paper, we use radiative transfer simulations of reionization to explore what measurements of the pkSZ power could tell us about the beginning of reionization and the ionizing emissions of the earliest galaxies observed by JWST. Another important (and related) question concerns the clustering of the ionizing sources. From an empirical perspective, essentially nothing is known about the source-halo connection at this epoch. The pkSZ power is in principle sensitive to the morphology of reionization, which is shaped by the clustering of its sources. It may thus contain precious information about the dark matter halos in which early ionizing sources resided. In this paper, we take up the question of whether pkSZ measurements could inform us about the galaxy-halo connection of the ionizing sources that began reionization.  As stated above, a key focus of this paper is to incorporate the most recent developments in our understanding of reionization's last phases into the discussion of how to interpret the pkSZ signal. As we will show, pinning down the end of reionization with Ly$\alpha$ forest measurements improves our ability to draw conclusions about the start of reionization.      

The remainder of this paper is organized as follows. In \S \ref{sec:Methods}, we describe our simulation methods. In \S \ref{sec:kSZhighzlim}, we explore using the pkSZ power to probe the starting epoch of reionization and the high-$z$ ionizing emissivity.  In \S \ref{sec:clustering}, we consider what the power can tell us about the clustering of the earliest ionizing sources. In \S \ref{sec:discussion}, we discuss the dependence of our results on modeling assumptions and compare our results to other works in the literature.  We offer concluding remarks in \S \ref{sec:conclusion}. In all calculations, we adopt a flat $\Lambda$CDM cosmology with $\Omega_m$=0.305, $\Omega_b = 0.048226$, H$_0$ = 68 km/s/Mpc, $\sigma_8$ = 0.82033, and $n_s$=0.9667.

\section{Methods}
\label{sec:Methods}

\subsection{Radiative Transfer Simulations of Reionization}
\label{sec:RTsims}

We have run a suite of radiative transfer (RT) simulations with the \textsc{FlexRT} code of Ref. \cite{Cain2024-ba}. We refer the reader to their paper for a detailed description of the main RT engine. Briefly, the code employs an adaptive ray tracing algorithm based on the Healpix RT formalism of Ref. \cite{2002MNRAS.330L..53A}, and a ray merging scheme akin to that of Ref. \cite{Trac2007-sb} for increased speed.  \textsc{FlexRT} adds to this a detailed sub-grid model for the LyC opacity of the IGM, described in Refs. \cite{Cain2021-yt, Cain2023-lm}, as well as the reionization I-front heating model of Ref. \cite{DAloisio2019}. The RT was run in post-processing on the cosmological hydrodynamics simulation from Ref. \cite{DAloisio2018}, which was run with the Eulerian code of Ref. \cite{Trac2004-ra}. The box size is $L_{\rm box} = 200~h^{-1}$ cMpc and the simulation contains $2 \times 2048^3 $ dark matter and gas resolution elements.   Ref. \cite{2024MNRAS.531.1951C} recently improved the halo mass resolution over that of Ref. \cite{DAloisio2018} by running with the same initial conditions an N-body-only simulation using $N = 3600^3$ particles.  Dark matter halos were identified using the spherical over-density method described in \cite{2015ApJ...813...54T}.  We find that the halo mass function in our simulation agrees with the halo mass function of Ref. \cite{2015ApJ...813...54T} to within 10 \%  down to a halo mass of $3\times 10^9~h^{-1}$ M$_\odot$.  However, in order to maximize the number of sources at $z>10$ in our simulations, we use all halos down to a mass of $10^9 h^{-1}$ M$_\odot$. At this mass, the simulated halo mass functions are suppressed by as much as $\sim 50\%$ with respect to the fits in Ref. \cite{2015ApJ...813...54T}.   We note that: (1) in reality, the cutoff in the ionizing sources induced by feedback/inefficient star formation is not perfectly sharp. In some sense, the incompleteness of the mass function crudely mimics a smoother decline; (2) the halos near this threshold are subdominant in their contribution to the total ionizing photon budget in our most realistic source models (c.f. Fig. 1 of Ref. \cite{Gangolli2024-dn}). 
  
We populate the dark matter halos with ionizing sources by first abundance matching to the observed rest-frame UV luminosity function. We use the luminosity function measurements of Ref. \cite{2021AJ....162...47B} at $z<8$, Ref. \cite{Adams2024-xg} at $8 < z < 14$, and Ref. \cite{Donnan2024-mw} at $z=14.5$. We extrapolate the luminosity function of \cite{Donnan2024-mw} out to $z=18$, the starting epoch of our simulations.  In the ensuing sections we describe our source modeling in detail, i.e. how exactly we assign ionizing emissivities to halos. The RT is performed on a uniform grid of size $N_{\rm RT}=200^{3}$.  For computational speed, we adopt the monochromatic approximation with photon energies $E_{\gamma} = 19$ eV. This energy is chosen to yield the same frequency-averaged hydrogen photoionization cross section as a power-law spectrum $I_{\nu} \propto \nu^{-1.5}$ between 1 and 4 Ry, where $I_{\nu}$ is the specific intensity.  This power-law spectrum approximates calculations of the time-averaged spectrum of metal poor stellar populations in synthesis models \cite[e.g.][]{DAloisio2019}.  We run the RT from $z=18$ down to $z=4.8$.

A key aspect of this work is the empirical calibration of our simulations to the observed Ly$\alpha$ forest mean transmission at 5 $\lesssim z \lesssim$ 6.  For this task we simulate the Ly$\alpha$ forest along 4,000 skewers traced at random directions from random locations in our hydrodynamic simulation volume (with $L_{\rm box} = 200~h^{-1}$Mpc and $N=2048^3$ gas resolution elements).  The gas properties on these skewers are rescaled to the RT grid values under the assumption of photoionization equilibrium, which is almost always a good approximation in the reionized IGM owing to the short photoionization equilibration timescale of the hydrogen gas.  For computing the Ly$\alpha$ optical depth, we employ the widely used approximation to the Voigt profile of Ref. \cite{2006MNRAS.369.2025T}. We apply two resolution corrections to the simulated mean flux: (1) the convergence correction described in the appendix of Ref. \cite{DAloisio2018}, which approximately corrects for spurious opacity in voids owing to the finite resolution of the hydrodynamic simulation; (2) the temperature-density correction described in the appendix of Ref. \cite{2024MNRAS.531.1951C},  which assigns a sub-grid temperature-density relation to each RT cell based on its local reionization redshift. This corrects for the relatively large cell sizes of our RT grid compared to the higher resolution of the hydrodynamic simulation. We refer the reader to those papers for detailed descriptions.   

To calibrate each simulation, we adjust the global ionizing emissivity -- particularly at $z\lesssim 7$ -- by trial and error until the simulated mean Ly$\alpha$ forest transmission agrees with observational measurements. This is equivalent to adjusting $\langle f_{\rm esc} \xi_{\rm ion} \rangle$ as a function of redshift at $z\lesssim 7$. (Plots illustrating this procedure, including the empirical constraints used, will be presented below.) This procedure would be extremely time consuming for the dozens of models explored in this paper, as each model requires at least several trial runs.  The reduced speed of light approximation is a commonly used method to reduce computational cost.  However, previous works have shown that it can lead to significant inaccuracies in the intensity and structure of the ionizing radiation background, particularly near the end of reionization \cite{2016ApJ...833...66G, 2019A&A...622A.142D, 2024MNRAS.531.1951C}.  Fortunately, Ref. \cite{Cain2024-ws} developed a rescaling/correction procedure that allows us to use the reduced speed of light approximation without sacrificing accuracy for the quantities of interest, greatly increasing the speed of the calibration and making this work tractable.  The following equation, motivated and tested by \cite{Cain2024-ws}, relates the global ionizing emissivity in a simulation with full speed of light, $\dot{N}_{\rm ion}^{c}$, to that in a simulation using reduced speed of light, $\dot{N}_{\rm ion}^{\Tilde{c}}$,  

\begin{equation}
    \label{eq:RSLA}
    \dot{N}_{\rm ion}^{c}(t) = \dot{N}_{\rm ion}^{\Tilde{c}} 
 - \dot{N}_{\gamma}^{\Tilde{c}} \left( 1 - \frac{\Tilde{c}}{c} \right),
\end{equation}
where $\Tilde{c}/c < 1$ is the ratio of the reduced speed of light to the speed of light, $c$. The quantity $N^{\Tilde{c}}_{\rm \gamma}$ is the mean number density of ionizing photons in the simulation with reduced speed of light, i.e.  

\begin{equation}
\label{eq:RSLA2}
N^{\Tilde{c}}_{\rm \gamma} = \int_0^t{dt \left[ \dot{N}^{\Tilde{c}}_{\rm ion} - \dot{N}^{\Tilde{c}}_{\rm abs}\right]},
\end{equation} 
where $\dot{N}^{\Tilde{c}}_{\rm abs}$ is the total absorption rate per unit volume. Given a simulation with reduced speed of light $\Tilde{c}$, we can use equations (\ref{eq:RSLA}) and (\ref{eq:RSLA2}) to map the quantities $\dot{N}_{\rm ion}^{\Tilde{c}}(t)$ and $N^{\Tilde{c}}_{\rm \gamma}(t)$ to $\dot{N}_{\rm ion}^{c}(t)$, giving approximately the emissivity history that would be needed in a simulation with full speed of light to produce the same reionization history, morphology, and key observables.  In practice, Ref. \cite{Cain2024-ws} showed that these equations recover the large-scale morphology and Ly$\alpha$ forest mean flux of a full-speed-of-light simulation to within $20\%$ for simulations with $\Tilde{c}/c \geq 0.2$ (the accuracy increases substantially with increasing $\Tilde{c}/c$); the global reionization history is recovered to within $1\%$ for $\Tilde{c}/c$ as low as 0.05.  In what follows, all of our simulations use $\Tilde{c}/c=0.2$.   We employ equations (\ref{eq:RSLA}) and (\ref{eq:RSLA2}) to map our emissivities to what they would be in a simulation with full speed of light.  All emissivities plotted in this paper correspond to $\dot{N}_{\rm ion}^{c}$.  We further test this rescaling procedure for the application at hand in Appendix \ref{append:rsla_vs_fsl}.  We compare the pkSZ power between full-speed-of-light-runs and reduced-speed-of-light-runs with $\Tilde{c}/c=0.2$.  We find that the above procedure recovers the pkSZ power spectra from full-speed-of-light runs to within $1.5 \%$, validating our approach.  We emphasize that the dozens of calibrated RT simulations run for this paper would have been a factor of $\approx 5\times$ more expensive without this time-saving (yet accurate) method.

\subsection{The kinetic Sunyaev-Zel'dovich Angular Power Spectrum}
\label{sec:kSZpow}
The temperature anisotropy induced by the kSZ effect can be expressed as the line-of-sight integral \cite{Sunyaev1980-pq},

\begin{equation}
    \label{eq:dtq}
    \frac{\Delta T}{T} = - \sigma_T \overline{n}_{e,0} \int e^{-\tau_{\rm es}} \frac{\hat{\gamma}\cdot\vec{q}}{c} \frac{ds}{a^2}, 
\end{equation}
where $\hat{\gamma}$ is a unit vector along the line of sight, $\sigma_T$ is the Thompson scattering cross section, $\overline{n}_{e,0}$ is the co-moving mean electron density, $a$ is the cosmological scale factor, and $s$ is the co-moving distance. The central quantity is the ionized momentum, $\vec{q}$ = (1 + $\delta) \chi \vec{v}$, where $\delta$ is the matter density contrast (with zero mean), $\chi = n_e / (n_H + 2n_{He}) $ is the free election fraction, and $\vec{v}$ is the peculiar velocity.  In this work we assume that the first reionization of helium traces hydrogen reionization and the second reionization of helium occurs instantly at $z$=3, although our results are insensitive to this choice. We compute the pkSZ angular power spectrum using the method of Ref. \cite{Park2013-bc}, which expresses the power as 

\begin{equation}
    \label{eq:Clpq}
    C_{\ell} = \left( \frac{\sigma_T \overline{n}_{e,0}}{c} \right)^2 \int \frac{ds}{s^2 a^4} e^{-2\tau} \frac{P_{q_{\perp}} \left(k = \ell/s, s\right)}{2},   
\end{equation}
where $ (2\pi )^3 P_{q_{\perp}}(k) \delta^{D}(\vec{k}-\vec{k'})=   \langle{\widetilde{q}_{\perp}(\vec{k}) 
  \cdot \widetilde{q}_{\perp}^* (\vec{k'})}  \rangle  $ is the power spectrum of the ionized momentum modes perpendicular to the Fourier wave vector, $\widetilde{q}_{\perp} = \widetilde{q} - \hat{k} (\widetilde{q} \cdot \hat{k})$ (the so-called ``transverse" modes), and $\delta^{D}(\vec{k}-\vec{k'})$ is the Dirac-delta function.  Note that $\widetilde{q} = \int \vec{q} e^{-i \vec{k} \cdot \vec{q}} d^3 \vec{x} $ denotes the Fourier transform of $\vec{q}$.  Only the $\widetilde{q}_{\perp}$ modes contribute significantly to the kSZ signal owing to the $\widetilde{q}_{||}$ contributions canceling out in the line-of-sight integral.  (See Refs. \cite{Ma2002,Park2013-bc} for a discussion.)  
  
In general, it is common to write the angular power spectrum in the dimensionless form: $D_{\ell} = C_{\ell} ( \ell + 1 ) \ell / (2 \pi)$.  For the most part, we will focus on the pkSZ power at $\ell=3000$ ($D_{3000}^{\rm pkSZ}$).  This is the angular scale of existing measurements due to the primary CMB strongly dominating at $\ell \lesssim$ 3000, and other foregrounds becoming increasingly problematic at $\ell \gtrsim$ 3000.  Ref. \cite{Jain2023-tk} forecasts that future experiments such as the Simons Observatory and CMB-S4 may be able to extend kSZ measurements to multiple bins in $\ell \in [2500, 5000]$. As we will show in \S \ref{sec:clustering}, the $\ell$-dependence of the power contains information about the clustering of reionization sources, so probing the power at several $\ell$ could be highly informative.   

Running RT simulations imposes practical limitations on the volume of our box. With our modest box sizes of $L_{\rm box} = 200~h^{-1}$cMpc, we miss large-scale velocity modes that are known to source kSZ power. To compensate for this, we add the correction term to $P_{q_{\perp}}$ developed by Ref. \cite{Park2013-bc},

\begin{equation}
     P_{q_{\perp}}^{{\rm miss}}(k,z) = {\int_{k<k_{{\rm box}}}} \frac{d^3 k'}{(2\pi)^3} (1-\mu^2) P_{\chi(1+\delta)}( | \vec{k} - \vec{k'} | ) P_{vv}^{{\rm lin}} ( k')
     \label{eq:missingpower}
\end{equation}
where $k_{\text{box}} = 2 \pi / L_{\text{box}}$ is the box-scale wave number, $\mu = \hat{k} \cdot \hat{k'}$, $P_{\chi(1+\delta)}(k)$ is the ionized matter power spectrum and $P_{vv}^{\text{lin}}(k)$ is the linear velocity power spectrum, which we evaluate using CAMB \cite{Lewis2000-vx}.   Ref. \cite{Alvarez2016} found that equation (\ref{eq:missingpower}) may nonetheless lead to an underestimate of $D^{\text{pkSZ}}_{3000}$ by $\sim$10-20$\%$ because the non-Gaussianity of reionization renders the irreducible component of $\langle \delta_{\chi} v \cdot \delta_{\chi}v\rangle$ non-negligible.  This does not affect our broad conclusions, however.  

The above integrals can be broken up into two contributions: (1) the so-called homogeneous kinetic Sunyaev-Zel'dovich Effect (hkSZ) from the post-reionization IGM; (2) the patchy kinetic Sunyaev-Zel'dovich Effect (pkSZ) sourced by reionization -- the quantity of interest here. Measurements of the pkSZ component require subtracting the hkSZ power.  For simplicity, and because reionization ends as late as $z_{\rm end} \approx5$ in some of our models, we define the pkSZ power to be the contribution from $z>5$, even for models in which reionization ends significantly earlier than $z=5$.  We note that the measurement of Ref. \cite{Reichardt2020-gp}, against which we will compare our models, adopts a slightly different demarcation of $z_{\rm end} = 5.5$. Using the scaling relationships in Ref. \cite{Shaw2012-md}, we find that rescaling the measurement to $z_{\rm end} = 5$ would increase the inferred pkSZ power by $\sim$ 0.1 $\mu$K$^2$ (by decreasing the hkSZ power for a fixed total kSZ). This is a relatively small effect that would not change our conclusions.

\section{Sensitivity of pkSZ Power to the Beginning of Reionization}
\label{sec:kSZhighzlim}

\subsection{Models}
\label{sec:models}

\begin{figure}
    \centering
    \includegraphics[scale=0.42]{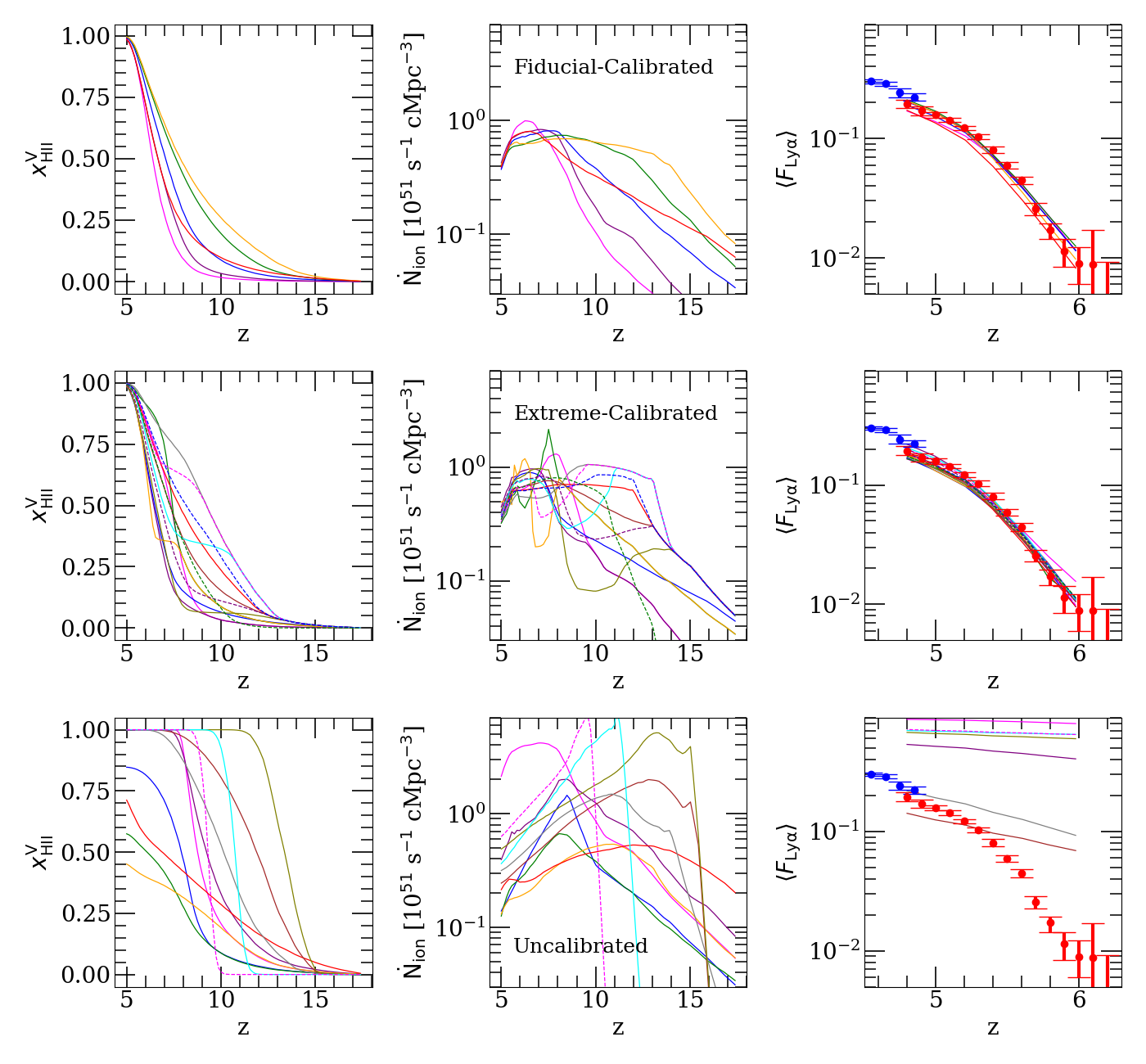}
    \caption{Summary of reionization models used in \S \ref{sec:kSZhighzlim}: the volume-weighted ionized fraction (left column), ionizing emissivity histories (center column), and forest transmission fraction (right column). The emissivities of the calibrated models are tuned for agreement with the observational measurements of Refs. \cite{Bosman2021}, shown as the red data points in the right column.  For reference, the blue data points show the lower redshift measurements of Ref. \cite{2013MNRAS.436.1023B}.  \textsc{Fiducial-Calibrated} models (top row) complete reionization in a manner consistent with the forest transmission fraction constraints, and have smooth/physical emissivity histories. The \textsc{Extreme-Calibrated} models have un-physically rapid and/or non-monotonic evolution in the emissivity. \textsc{Uncalibrated} models (bottom row) complete reionization far too early or late to agree with the Ly$\alpha$ forest constraints.  The \textsc{Extreme-Calibrated} and \textsc{Uncalibrated} models are for illustrative purposes only; they are used to test the generality of our main results below.}
    \label{fig:models}
\end{figure}

In this section, we adopt a source model in which the ionizing emissivity of a galaxy is assumed proportional to its rest-frame UV luminosity, $\dot{n}_{\rm ion} \propto L_{\rm UV}$. Since the UV-luminosity of a galaxy is uniquely mapped to a halo mass through abundance matching, we refer to this as the ``emissivity-halo'' prescription. The emissivity-halo prescription sets the clustering of the ionizing sources, which, in turn, shapes the morphology of reionization.   The current section focuses on what the pkSZ power can tell us about the ionizing output of high-$z$  galaxies {\it for a fixed emissivity-halo prescription.}  In \S \ref{sec:clustering}, we will explore how changing the model (thus changing morphology of reionization) affects our results. Note that, although we fix the emissivity-halo prescription, we still vary the overall normalization of the galaxy population's ionizing emissivity, i.e. the highly uncertain $\langle f_{\rm esc} \xi_{\rm ion}\rangle(z)$.  In this manner, we have run a suite of simulations spanning a wide range of emissivity histories. In the ensuing discussion, we break these models up into three categories. 

We begin with our fiducial models. The left and middle panels in the top row of Figure \ref{fig:models} show the reionization and emissivity histories, respectively, for a set of models that we calibrated to the Ly$\alpha$ forest mean transmission at $5 < z < 6$, as described in \S \ref{sec:RTsims}. The right panel compares the Ly$\alpha$ mean transmissions in the models to a collection of recent observational measurements.  These models span a range of $\tau_{\rm es} \in [ 0.047, 0.068 ]$, compatible with the most recent measurement of $\tau_{\rm es} = 0.058\pm 0.006$ based on Planck data \cite{Tristram2024}.  We term these models \textsc{Fiducial-Calibrated}.


In the next section, we will find that the starting epochs and cumulative ionizing photon outputs of the \textsc{Fiducial-Calibrated} models scale with the kSZ power, suggesting that we may use existing limits on the power to draw broad conclusions about the high-$z$ galaxy population.  As a foil to these models, we have constructed a set of extreme scenarios.  These are also calibrated to the forest transmission, but with emissivity histories designed to test the generality of the scaling relations. The middle row of Figure \ref{fig:models} shows our \textsc{Extreme-Calibrated} models.  The main features that result in our labeling these models ``extreme'' are the rapidity of the emissivity evolutions and their non-monotonic nature. In these models, the time-scale for the evolution, $t_{\rm{em}} = \dot{N}_{\rm ion} (\frac{d \dot{N}_{\rm ion}}{dt})^{-1}$, is at times much shorter than ($<  1/10$th) the Hubble time. In contrast, the \textsc{Fiducial-Calibrated} models have $t_{\rm{em}} = \dot{N}_{\rm ion} (\frac{d \dot{N}_{\rm ion}}{dt})^{-1} \sim H(z)^{-1}$ at all times before $z=6$.\footnote{This discussion pertains to $z>6$, before the abrupt drop in ionizing emissivity seen at $z\lesssim 6$, which has been noted by several other authors (see e.g. Ref. \cite{2024MNRAS.531.1951C} and references therein). The origin of this drop is still a matter of debate, which we do not take up here.  We note that the contribution to the pkSZ power below $z=6$ is small.} These models nonetheless latch on to the forest constraints at $z < 6$ by construction and have CMB optical depths spanning $\tau \in$ [0.048, 0.073]. We emphasize that the \textsc{Extreme-Calibrated} models are illustrative in nature, and are not intended to represent physically plausible reionization scenarios.  

Lastly, we consider another set of illustrative \textsc{Uncalibrated} models, shown in the bottom row, that are not calibrated to the forest constraints. As the bottom-left and bottom-right panels of Figure \ref{fig:models} show, reionization ends significantly earlier in these models than the forest constraints allow.  This also results in some fairly high CMB optical depths spanning $\tau_{\rm es} \in [0.051, 0.119]$.   The purpose of the \textsc{Uncalibrated} models is to show how the forest constraints, when combined with the kSZ power, can be used to sharpen our view of reionization's early phases.  

\subsection{Results}

\subsubsection{The Starting Epoch of Reionization}
\label{sec:zstart}
\begin{figure}
    \centering
    \includegraphics[width=0.9\linewidth]{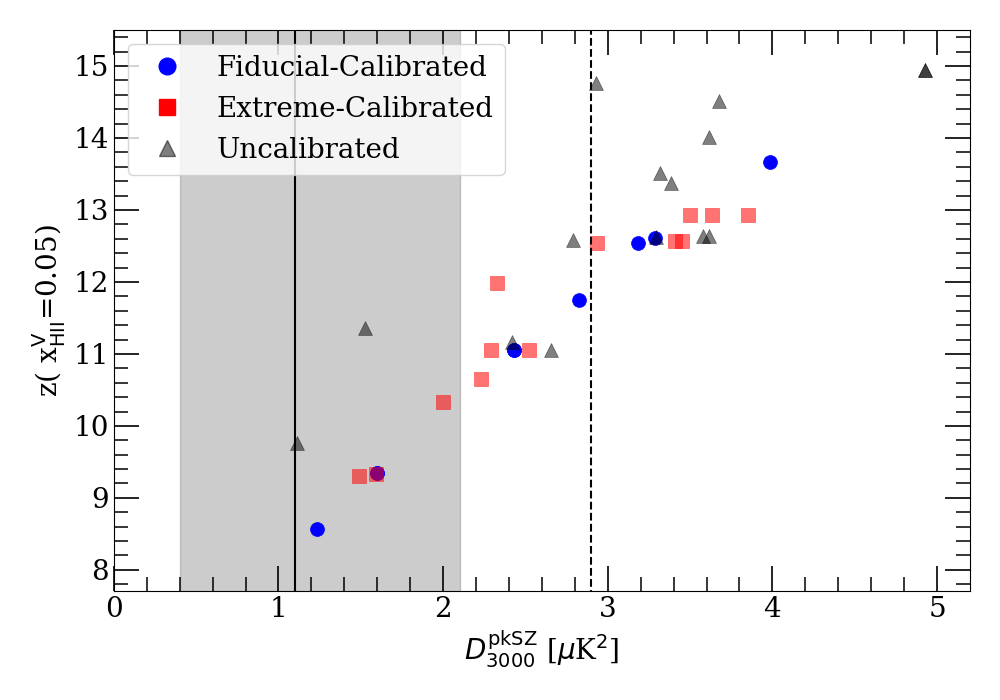}
    \caption{Correlation between patchy kSZ power at $\ell=3000$ and the starting epoch of reionization (defined to be the redshift at which $x_{\rm HII}^{V}$=$5\%$). All models shown here adopt the same model for the clustering of ionizing sources. The \textsc{Fiducial-Calibrated} models (blue circles) exhibit the tightest correlation between $D_{3000}^{\rm pkSZ}$ and $z_{05}$, but the correlation is still apparent for our other models. These results demonstrate that fixing the end of reionization, in accordance with Ly$\alpha$ forest constraints, makes the pkSZ power a probe of the starting epoch of reionization. In \S \ref{sec:clustering}, we explore the effects of varying the source clustering model. }
    \label{fig:z05vsD3k}
\end{figure}

\begin{figure}
    \centering
    \includegraphics[width=0.9\linewidth]{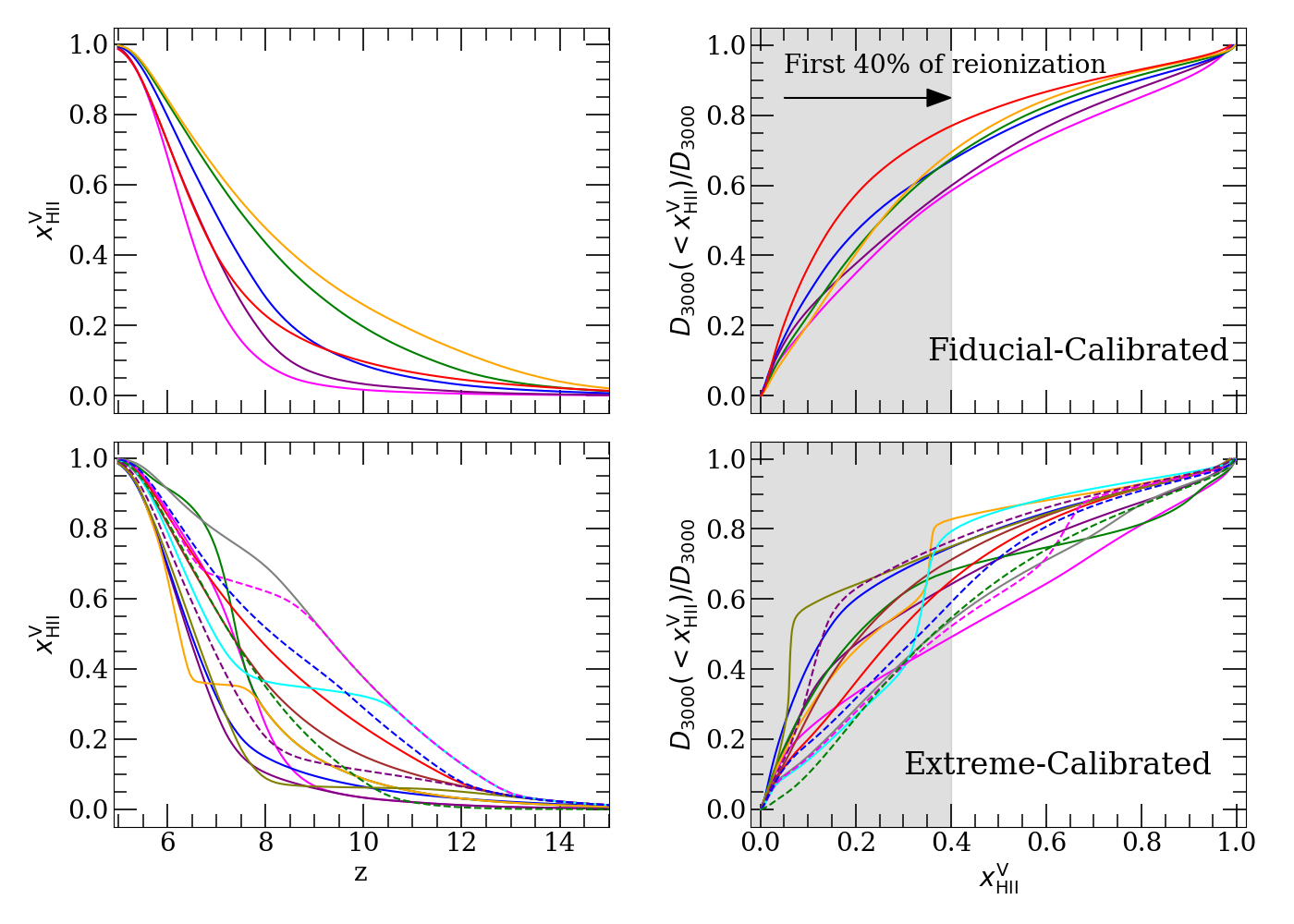}
    \caption{The accumulation of pkSZ power as a function of the global ionized fraction. The right panels show the fraction of $D_{3000}$ accumulated before $x^{\rm V}_{\rm HII}$, while the left panels show the corresponding reionization histories for reference.  The top and bottom rows show our \textsc{Fiducial-Calibrated} and \textsc{Extreme-Calibrated} models, respectively.  For all of these models, 50-85\% of $D_{3000}$ comes from the earlier phases of reionization, $x_{\rm HII} \le 40\%$, which we illustrate with the shaded gray region.       }
    \label{fig:Dl_vs_xandz}
\end{figure}

The first of our main results is displayed in Fig. \ref{fig:z05vsD3k}, where we show the pkSZ power versus the starting epoch of reionization, $z_{05}$, defined here as the redshift at which the volume-average ionized fraction reaches 5$\%$.  The data points correspond to our models, with different plot symbols denoting the categories introduced in the last section. For reference, the solid line, gray shaded region and black dashed line correspond to the measurement, 1-$\sigma$ error bars and 95\% confidence upper limit of Ref. \cite{Reichardt2020-gp}, respectively.   There is a tight correlation between $D_{3000}^{\rm pkSZ}$ and $z_{05}$, especially among the \textsc{Fiducial-Calibrated} models (blue circles), noting again that all of the models in this section adopt effectively the same source clustering model.  Qualitatively, the existence of a tight correlation is not sensitive to our choice for the ionized fraction threshold defining the starting epoch.   The origin of this correlation can be understood as follows. Previous studies have highlighted the dependence of the pkSZ power on reionization's duration and midpoint \cite{Battaglia2012, Chen2022, Nikolic2023, Alvarez2016, McQuinn2005-re}, although see Refs. \cite{ Mesinger2012-ad, Park2013-bc, Gorce2020-gx} and \S \ref{sec:discussion} for nuances.  For most of our models, the end of reionization is tightly constrained by the Ly$\alpha$ mean transmission. While there is some variation among our models in the midpoint, most lie within the range permitted by CMB $\tau_{\rm es}$ measurements.  With the endpoint and midpoint of reionization being roughly fixed in most of our models, and the source clustering being fixed, it follows that $D_{3000}^{\rm pkSZ}$ scales with the starting epoch.  

It is perhaps surprising that the \textsc{Extreme-Calibrated} models do not exhibit more scatter in the relationship between $z_{\rm 05}$ and $D_{3000}$, raising the broader question of why the correlation is so tight for the calibrated models. It is instructive to consider the conditions that give rise to the most extreme outlier, the \textsc{Uncalibrated} model denoted by the triangle at $D_{3000} \approx 3~\mu$K$^2$, $z_{05}$=14.8. The reionization history is quick and early in this model, with $x_{\rm HII}^{\rm V} = 1$\% and 99\% at $z \approx 15.1$ and 11.3, respectively -- a time span of just 130 Myr. The reionization history corresponds to the right-most olive curve in the bottom-left panel of Fig. \ref{fig:models}.\footnote{Note that this model has a large pkSZ power in spite of its short duration because of our convention that all contributions from $z \ge$5.0 are included in our definition of the pkSZ signal. We emphasize, however, that the existence of the tight correlation in Fig. \ref{fig:z05vsD3k} does not depend on this definition; the \textsc{Fiducial-Calibrated} and \textsc{Extreme-Calibrated} models are constructed to end near $z = 5 - 5.5$, so the changes in power are insignificant if we instead chose to place the demarcation between pkSZ and hkSZ at $z_{99}$.} This scenario is not only un-physical in how quickly the ionizing emissivity rises; it also badly violates the Ly$\alpha$ forest constraints for the end of reionization. More broadly, for a fixed $z_{05}$, we find that it is difficult to construct a model with a much smaller $D_{3000}$ that scatters far to the left, i.e. significantly above the relation, without violating the Ly$\alpha$ forest constraints imposed at $z\lesssim 6$.  

What about scattering below the relation? Again, we find that it is generally difficult to construct models that increase the pkSZ power at a fixed $z_{05}$.  To help understand why, in Fig. \ref{fig:Dl_vs_xandz} we plot the fraction of $D_{3000}^{\rm pkSZ}$ accumulated at times before a global ionized faction of $x^{\rm V}_{\rm HII}$ (right panels).  For reference, the corresponding reionization histories are shown in the left panels. The top and bottom panels correspond to our \textsc{Fiducial-Calibrated} and \textsc{Extreme-Calibrated} models, respectively.  These panels show that $D_{3000}$ receives most (50-85\%) of its contribution from the earlier phases of reionization, $x^{\rm V}_{\rm HII} < 40$\%.  To construct a model that maximizes $D_{3000}$ for fixed $z_{05}$ (i.e. scatters significantly below the relation), features in the ionization fields with angular scale $\ell\sim3000$, corresponding to $\theta \sim 0.06$ degrees, must be present for more of the light cone.  To this end, we consider two models where the global ionized fraction lingers near $\sim 30\%$ for much of reionization's duration. These are the two models in the bottom-right panel with sharp rises in $D_{\rm 3000}$ at $x^{V}_{\rm HII}\approx 30\%$.  Even in these extreme scenarios, designed to pump up $D_{\rm 3000}$, we find that $D_{3000}$ increases by a modest $\sim 15\%$. Again, further imposing conditions on the ionizing emissivity evolution based on physical grounds limits the scatter even more, as indicated by the blue circles in Figure \ref{fig:z05vsD3k}.  
 
From Fig. \ref{fig:z05vsD3k} we are led to several broad conclusions that apply for a fixed source clustering model. Reionization models that are calibrated to the Ly$\alpha$ forest mean transmission measurements at $z\lesssim 6$ generally exhibit a correlation between $D_{3000}^{\rm pkSZ}$ and the starting epoch of reionization, quantified here with $z_{05}$.  This matches expectations from the literature, with the notable exception of Ref. \cite{Park2013-bc}, which found $D_{3000}$ was not sensitive to $z_{05}$.  We discuss this discrepancy more in \S \ref{sec:discussion}.  The correlation we find is tighter if the models are restricted to those with physically reasonable ionizing emissivity histories, i.e. monotonically rising at a rate of order the Hubble time, as in our \textsc{Fiducial-Calibrated} models. Based on these findings, Figure \ref{fig:z05vsD3k} indicates that the existing 95\% upper-limit on the pkSZ power reported by SPT \cite{Reichardt2020-gp} implies a corresponding rough upper limit of $z_{\rm 05} \lesssim 12$ for our $\dot{n}_{\rm ion} \propto L_{\rm UV}$ model. For the ACT measurement of Ref. \cite{Beringue2025-ka}, we adopt a conservative (low) estimate for the hkSZ contribution of 0.85 $\mu$K$^2$, based on Ref. \cite{Shaw2012}, to obtain an approximate 2 $\sigma$ upper limit of $D^{\rm pkSZ}_{3000} \lesssim 2.05$ $\mu$K$^2$. This translates to $z_{05}\lesssim 11$ for the ACT measurement.  These results are consistent with $z_{05}$ values implied by constraints on the mid-point and duration of reionization in other works using the pkSZ \cite{Nikolic2023,Chen2022,Choudhury2020-gy}.  

Figure \ref{fig:z05vsD3k} highlights how the sharpening of pkSZ measurements by current and forthcoming CMB surveys, when combined with Ly$\alpha$ forest measurements, could place more stringent limits on the starting epoch of reionization. We emphasize that different assumptions about the source clustering will in general add variations to the pkSZ power for a fixed reionization history, a topic we explore in \S \ref{sec:clustering}.  Before addressing that, we turn to the ionizing photon output of the galaxy population. 

\subsubsection{Ionizing Photon Production of Early Galaxies}

\begin{figure}
    \centering
    \includegraphics[width=1\linewidth]{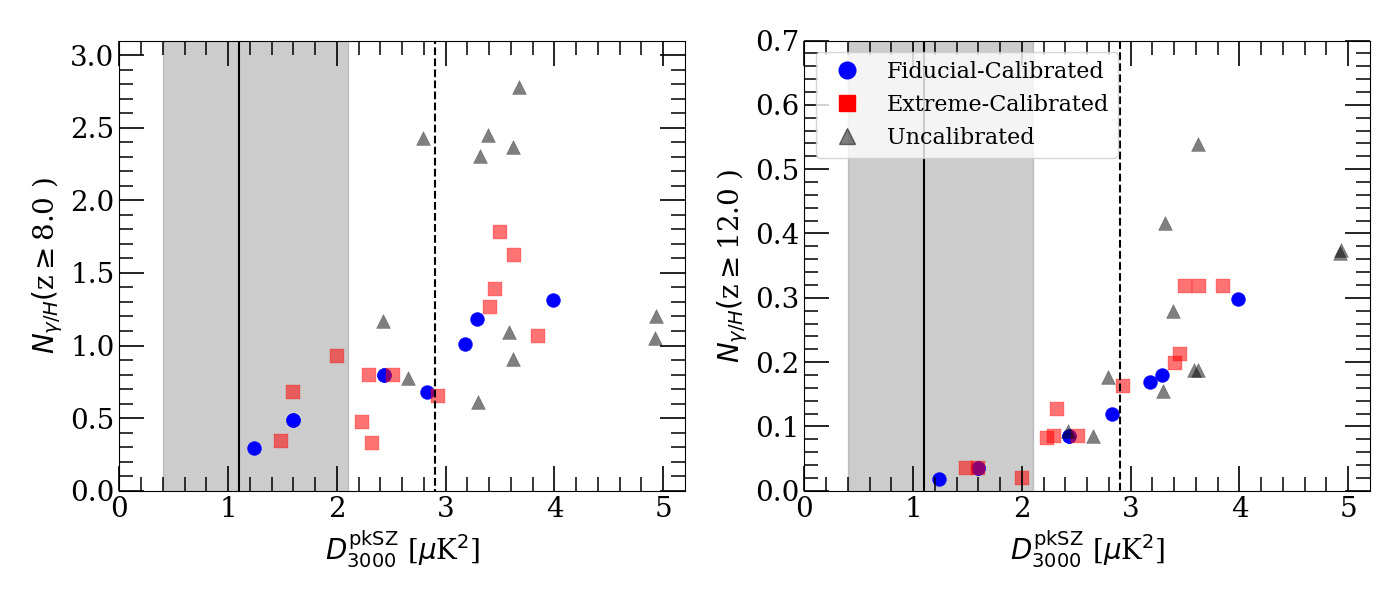}
    \caption{Correlation of the ionizing photon budget of the galaxy population with the pkSZ power given Ly$\alpha$ forest constraints on the end of reionization.  The left and right panels show the cumulative number of ionizing photons emitted into the IGM per hydrogen atom before a redshift of $z_*=8$ and $z_*=12$, respectively (see eq. \ref{eq:NgammaH}) as a function of pkSZ power.  In both  panels, there is a prominent trend between the $N_{\gamma /H}^{z_*}$ and $D_{3000}$, especially for the \textsc{Fiducial-Calibrated} models. Existing $D_{3000}$ measurements can place model-dependent upper bounds on the ionizing photon outputs of early galaxies.  }
    \label{fig:NgH}
\end{figure}

The starting epoch of reionization is set by the ionizing photon output of the high-$z$ galaxy population. We now explore what the pkSZ power can tell us about this output.  We quantify it with the the cumulative number of ionizing photons emitted into the IGM per hydrogen atom up to a cosmic time $t_*$:

\begin{equation}
    \label{eq:NgammaH}
    N_{\gamma / H}(z \ge z_*) = \int_{t_0}^{t_*} \frac{\dot{n}_{\rm ion}(t)}{n_{\rm H}(t)} dt,
\end{equation}
where $\dot{n}_{\rm ion}$ is the global rate of ionizing photon emission per unit volume, $n_{\rm H}$ is the cosmic mean hydrogen density, and $z_*$ is the redshift at the reference time $t_*$. In what follows, we will use the notation $N_{\gamma/H}^{z_*} \equiv N_{\gamma / H}(z \ge z_*)$.  We begin the integration at $z=18$, consistent with the start of our RT simulations.  

In Figure \ref{fig:NgH} we plot $N_{\gamma/H}^{z_*}$ versus $D_{3000}$ for $z_*=8$ and $z_* =12$ in the left and right panels, respectively.  For reference, the cosmic time interval between these redshifts is $\approx 270$ Myr.  Again, the solid black line, shaded gray region and dashed line denote the mean value, 1-$\sigma$ error bars and 95$\%$ confidence upper limits of the SPT measurement \cite{Reichardt2020-gp}.  A key result here is the correlation between $D_{3000}$ and $N_{\gamma / H}^{z*}$ in the \textsc{Fiducial-Calibrated} models (blue-circles).  This result is a consequence of the $z_{\rm 05}$ - $D_{3000}$ correlation discussed in the previous section, and the fact that the reionization history is set by the ionizing photon production of the sources minus the absorption rate by the sinks.  Note that there is significantly less scatter in the $N_{\rm \gamma/\mathrm{H}}^{12}-D_{3000}$ relation (right panel), compared to the $N_{\rm \gamma/\mathrm{H}}^8-D_{3000}$ relation (left panel) -- particularly for $D_{3000} \lesssim 3.0 \mu$K$^2$.  Appealing to Figure \ref{fig:z05vsD3k}, we can see that models with $D_{3000} \lesssim 3.0 \mu$K$^2$ should have $z_{05} \lesssim 13$.  Hence, any variation in $N_{\gamma/H}^{12}$ for fixed $D_{3000}$ must arise from variations in the reionization history within a quite narrow window range in cosmic time, which reduces the scatter among these models. 

Figure \ref{fig:NgH} suggests that we can estimate a rough upper limit on the ionizing photon production of early galaxies in our models.  The 95\% upper SPT limit on $D_{3000}$ translates to a limit of $N_{\gamma/\mathrm{H}} \lesssim 1.0$ ionizing photon produced per hydrogen at $z>8$ for our $\dot{n}_{\rm ion}\propto L_{\rm UV}$ source model.  This becomes $N_{\gamma/\mathrm{H}} \lesssim 0.15$ for $z>12$. Regarding the ACT measurement, if we again assume a value of 0.85 $\mu$K$^2$ for the hkSZ contribution, the lower $D_{3000}$ reported by them tightens these limits to $N_{\gamma/H} \lesssim 0.7$ and 0.08 for $z \geq 8$ and 12, respectively. (All estimates here do not include our \textsc{extreme} and \textsc{uncalibrated} models.)  It is interesting to compare these estimates to the models in Ref. \cite{2024MNRAS.535L..37M}, which were mentioned in the introduction to this paper. In their fiducial ``photon budget crisis'' model, the observed galaxy population emits $\approx 1.6$ photons per hydrogen atom into the IGM before $z=8$. This number increases to $\approx 3.1$ if the UV luminosity function is extrapolated down to $M_{\rm UV} = -13$. 

The results in this section, which are restricted to a single emissivity-halo prescription, have served mainly to illustrate the relationships between the reionization history, ionizing emissivity, and pkSZ power, and to build intuition for them.  We now turn to the important effect that source clustering has on these relationships. 

\section{Effect of Source Clustering on pkSZ Power}
\label{sec:clustering}

\subsection{Quantifying the Effect of Source Clustering}
 
In this section, we investigate how the pkSZ power depends on the assumed model of source clustering.  As discussed in the introduction, the faint end of the UV luminosity function during reionization remains highly uncertain.  Moreover, there is no broad consensus on the ionizing photon production efficiencies and escape fractions of even the observed high-$z$ galaxies. As such, we adopt 3 emissivity-halo prescriptions that are designed to bracket a wide range of source clustering possibilities:

\begin{itemize}

    \item \textsc{$\dot{n}_{\rm ion} \propto L_{\rm UV}$} -- This is the same model used in previous sections.  Recall that all halos more massive than than $10^{9}~h^{-1}$M$_{\odot}$ are sources of ionizing photons. The rate of ionizing photon emission into the IGM is assumed proportional to UV luminosity.  

    \item \textsc{UV-Bright} -- This model also assumes the proportionality $\dot{n}_{\rm ion} \propto L_{\rm UV}$, but implements a cutoff such that only galaxies brighter than $M_{\rm UV} = -18$ contribute.  This is a scenario in which the brightest observed galaxies are the sole drivers of reionization (see e.g. \cite{Naidu2020,Matthee2021-ip} for empirical arguments in favor of related scenarios).

    \item \textsc{Democratic} - In this model, all galaxies emit ionizing photons at the same rate, independent of UV luminosity ($\dot{n}_{\rm ion} \propto$ const).  Effectively, this represents a scenario in which the ionizing photon budget is dominated by the faintest (and therefore undetected) galaxies, owing to their abundance.
     
\end{itemize}

In the standard picture of hierarchical galaxy formation, it is generally assumed that scenarios like our \textsc{UV-Bright} model would result in a more delayed and perhaps rapid reionization process, compared e.g. to the \textsc{Democratic} model.  This assumption is rooted in the fact that the most massive galaxies, on average, form later in the cosmological timeline. But given the large uncertainties in the ionizing photon outputs of galaxies, here we attempt to decouple the effects of source clustering from the reionization history. This allows us to isolate the effects of source clustering/morphology on the pkSZ power. We note that this approach is different from many previous discussions of source clustering/morphology in the pkSZ literature \cite[e.g.][]{Battaglia2012, Park2013-bc, Mesinger2012-ad, Nikolic2023}, which do not attempt to isolate those effects from differences in the reionization history (but see \cite{Chen2022,Paul2020-ub} for studies that do).    

To this end, we select a few representative \textsc{Fiducial-Calibrated} reionization histories from the models of the last section, which adopt the $\dot{n}_{\rm ion} \propto L_{\rm UV}$ emissivity-halo prescription.  We run two variations of each model using the \textsc{UV-Bright} and \textsc{Democratic} emissivity-halo prescriptions.  Importantly, we do not change the input emissivity histories for these variations.  This results in small changes in the early global reionization history, $\Delta x_{\rm HII} \sim 0.01$ at $z\sim 8 -10$, and somewhat larger changes later, $\Delta x_{\rm HII} \sim 0.05$ at $z \sim 5.5-6.5$.  We note that these small changes should not significantly affect our results, as the majority of the $D_{3000}$ contributions come from early times, as shown above.  The variation in the source clustering has a larger impact on the pkSZ power through changes in the morphology of ionized regions -- the effect that we focus on in this section.\footnote{With this procedure, we also find that the calibration to the Ly$\alpha$ forest mean transmission fraction is mostly preserved; the deviations are again largest at $z\sim 6$, where they vary by $\sim 50$\%.  We do not recalibrate the models, as doing so would not significantly change $D_{3000}$.} 

\begin{figure}
    \centering
    \includegraphics[width=0.95\linewidth]{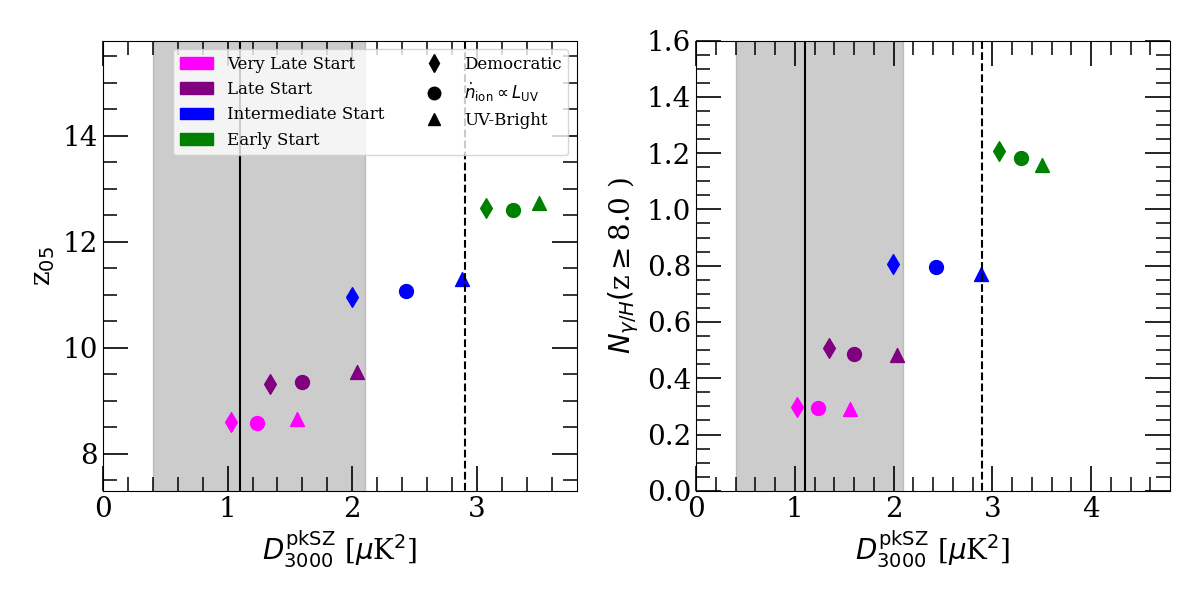}
    \caption{ The effect of source clustering/reionization morphology on the pkSZ power. The left and right panels show the $D_{3000}-z_{05}$ and $D_{3000}-N_{\gamma/H}^{8}$ relationships, respectively.   Colors correspond to different reionization histories, while marker styles correspond to different emissivity-halo prescriptions. For a fixed color, the range in $D^{\rm pkSZ}_{3000}$ values quantifies how much the pkSZ power varies under our different source prescriptions for a fixed reionization history.  Models with more biased ionizing sources exhibit more large-scale power in the ionization field which, in turn, leads to more pkSZ power. }
    \label{fig:source_z05_Ngh}
\end{figure}

\begin{figure}
    \centering
    \includegraphics[width=0.95\linewidth]{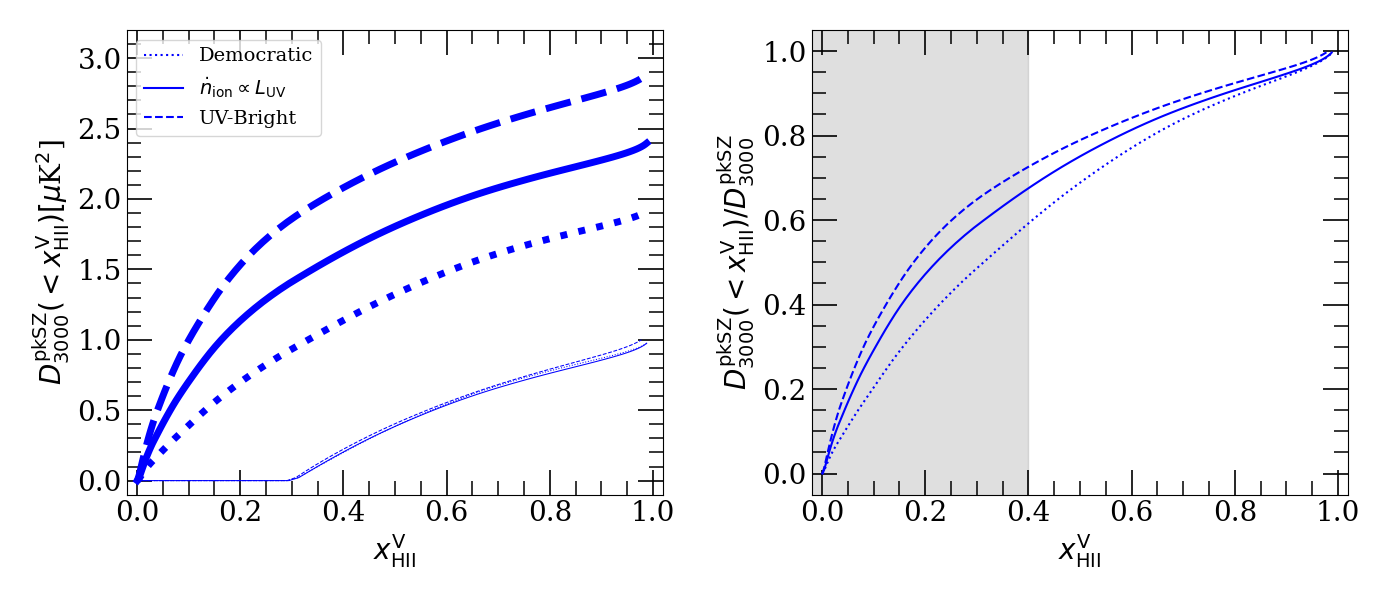}
    \caption{Source clustering imprints differences in the pkSZ power within the first $30$\% of reionization.  \textit{Left:} The accumulation of $D_{3000}^{\rm pkSZ}$ as a function of $x_{\rm HII}^{\rm V}$ compared among our emissivity-halo prescriptions for the fixed \textsc{Intermediate Start} reionization history (same as the blue points in Fig. \ref{fig:source_z05_Ngh}). Thick curves show the accumulation over the entire reionization history, whereas the thin curves show only the growth in $D^{\rm pkSZ}_{3000}$ over the last 70\% of reionization.    \textit{Right:} The fraction of $D_{3000}^{\rm pkSZ}$ accumulated by an ionized fraction of $x_{\rm HII}^{\rm V}$ for the same model set.  For a fixed reionization history, the fraction of power that originates from $x_{\rm HII}^{\rm V}\lesssim40\%$ scales with the clustering strength of the sources.   }
    \label{fig:source_cDl}
\end{figure}

Figure \ref{fig:source_z05_Ngh} quantifies the variations in $D_{3000}^{\rm pkSZ}$ that arise from our different source clustering prescriptions.  We show both $z_{05}$ vs $D_{3000}^{\rm pkSZ}$ (left panel) and $N_{\gamma/H}^8$ vs $D_{3000}^{\rm pkSZ}$ (right panel). The different colors denote different reionization histories, as labeled qualitatively in the plot legend according to when reionization starts.  The circles, triangles, and diamonds correspond to the \textsc{$\dot{n}_{\rm ion} \propto L_{\rm UV}$}, \textsc{UV-bright}, and \textsc{Democratic} emissivity-halo prescriptions, respectively. Note, for a given reionization time line (e.g. the \textsc{Intermediate start} models), the fact that $z_{05}$ and $N_{\gamma/H}^8$ stay approximately constant between the different source prescriptions indicates that the three reionization histories are approximately the same, as already noted above. (The small variations in $N^8_{\rm \gamma / \mathrm{H}}$ seen in the right panel owe to small inaccuracies in our RSLA correction.   Although we hold the $\tilde{c}/c = 0.2$ emissivities fixed, applying the RSLA correction for simulations with different source models can produce slightly different $\tilde{c}/c = 1$ results -- see Eq.~\ref{eq:RSLA}.) The corresponding range in the $D^{\rm pkSZ}_{3000}$ values quantifies how much the pkSZ power can vary under different source prescriptions for a fixed reionization history.   For example, for the \textsc{Intermediate Start} models, $D^{\rm pkSZ}_{3000}$ varies between $1.98$ $\mu$K$^2$ for the \textsc{Democratic} model to $2.88$ $\mu$K$^2$ for the \textsc{UV-bright} model. Indeed, we find variations of 35-50\% owing to source clustering alone across the different models.\footnote{This is a larger spread than seen by Ref. \cite{Chen2022}, but consistent with Refs. \cite{Paul2020-ub, Jain2023-tk}. Our models likely bracket a much wider range of reionization morphologies than in Ref. \cite{Chen2022}, which may account for the difference.}  Intuitively, models with more biased ionizing sources exhibit more large-scale power in the ionization field in the first $\sim$ third of reionization which, in turn, leads to more pkSZ power for a fixed reionization history.  

Figure \ref{fig:source_cDl} considers the question of where these differences arise. There we compare emissivity-halo prescriptions for a fixed reionization history (\textsc{Intermediate Start}). Similar to Figure \ref{fig:Dl_vs_xandz}, the right panel shows the fraction of $D^{\rm pkSZ}_{3000}$ accumulated before a time corresponding to $x^{\rm V}_{\rm HII}$.     In the left panel, the thick/top set of curves show the absolute $D^{\rm pkSZ}_{3000}(< x^{\rm V}_{\rm HII})$ over the entire reionization history, while the thin/bottom set of curves start the accumulation at $x^{\rm V}_{\rm HII} = 30$\%.  The results in the left panel make clear that the differences in $D^{\rm pkSZ}_{3000}$ across source models occurs almost exclusively at early times, $x_{\rm HII}^{\rm V}\lesssim30\%$.  In fact, for the reionization histories considered in this section, we find that the relative difference in power originating from the last $\sim70\%$ of reionization is $\lesssim5\%$ across source models.\footnote{In absolute terms, this reflects a typical difference of $\sim0.05 \mu$K$^2$, but goes up to $\sim0.1 \mu$K$^2$ for our \textsc{Early Start} model.}   

We emphasize that the results in this section should be taken as a rough upper limit on the effect of source clustering, for a fixed reionization history. If the source clustering is connected to the reionization history in the way usually assumed, it would decrease the spread in $D^{\rm pkSZ}_{3000}$ values. For example, if reionization is delayed in our \textsc{UV-Bright} model (because we have to wait for more massive halos to form), this would have the effect of decreasing $D^{\rm pkSZ}_{3000}$, therefore bringing it closer to the \textsc{$\dot{n}_{\rm ion} \propto L_{\rm UV}$} result. Likewise, $D^{\rm pkSZ}_{3000}$ would move upward in our \textsc{Democratic} model if reionization starts earlier in that scenario.  We will discuss scenarios with evolving emissivity-halo prescriptions in \S \ref{sec:evolvingsources}.

\begin{figure}
    \centering
    \includegraphics[width=0.95\linewidth]{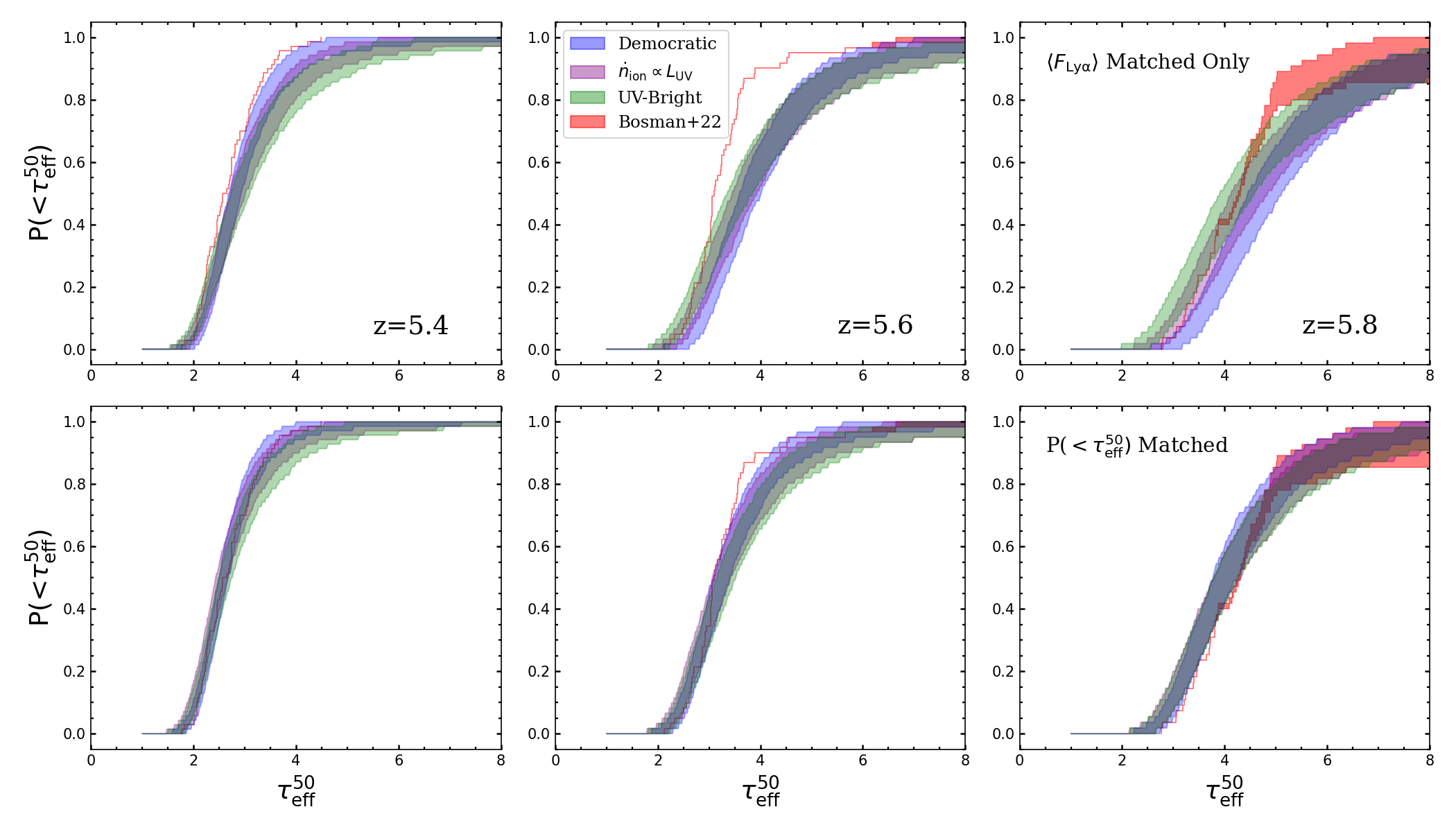}
    \caption{Adjusting the calibration of our models for agreement with the observed Ly$\alpha$ forest opacity fluctuations. The panels show the cumulative probability distribution function of the Ly$\alpha$ forest effective optical depth, $\tau_{\rm eff}$. The red histograms show observational measurements from Ref. \cite{Bosman2021}. The other colors correspond to different emissivity-halo prescriptions for a fixed reionization history (\textsc{Late Start}; purple data points in Fig. \ref{fig:source_z05_Ngh}). The top row shows distributions from our original models, which are calibrated only to the mean transmission fraction, $\langle F_{Ly\alpha}\rangle$. Despite this calibration, the models generally do not agree well with the observed distributions.  The bottom row shows the models after they have been recalibrated to the observed distributions. The left panel of Figure \ref{fig:Ptau_xHII} shows how the reionization histories were perturbed to achieve these recalibrations. }
    \label{fig:Ptau_cal}
\end{figure}

\subsection{Breaking the Degeneracy with Source Clustering}

The results in Figure \ref{fig:source_z05_Ngh} show how uncertainties in the sources of reionization complicate the connection between $D^{\rm pkSZ}_{3000}$ and the ionizing output of high-$z$ galaxies. The issue is that the effect of reionization's morphology on $D^{\rm pkSZ}_{3000}$ (as set by the source clustering) is degenerate with the reionization history. In this section, we explore the possibility of breaking this degeneracy with additional information. Section \ref{sec:taueff} considers the usage of the $z = 5-6 $ Ly$\alpha$ forest opacity fluctuations, while Section \ref{sec:pkSZshape} considers the use of the shape of the pkSZ spectrum.

\begin{figure}
    \centering
    \includegraphics[width=0.98\linewidth]{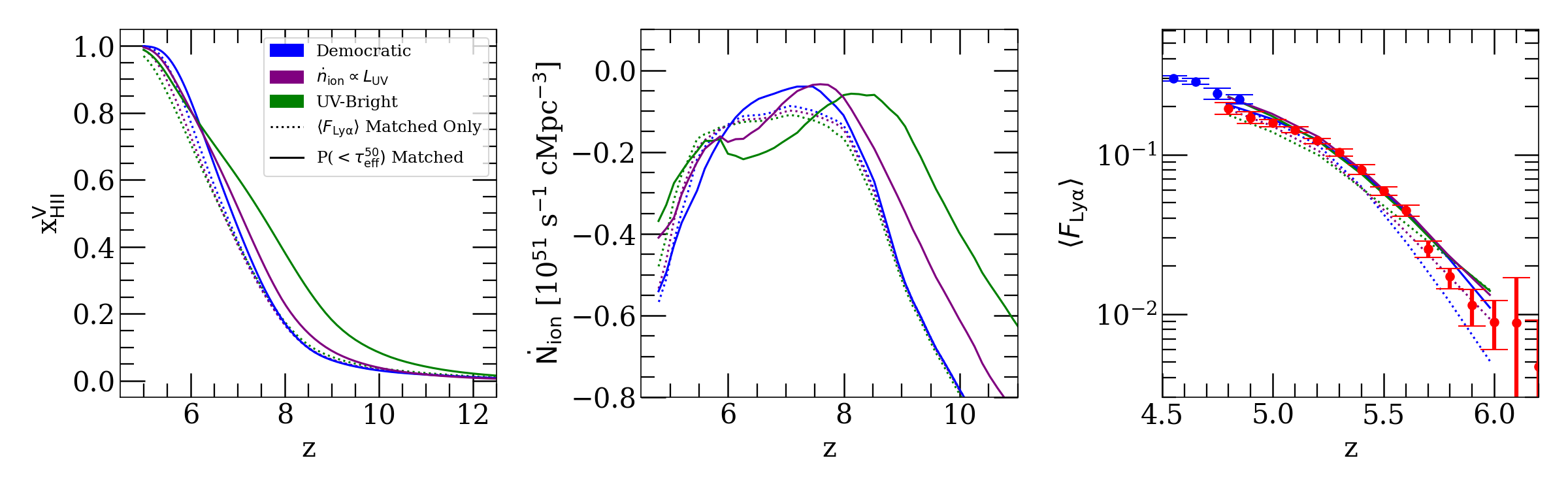}
    \caption{Recalibrating our models for agreement with the Ly$\alpha$ forest opacity fluctuations. The left, middle, and right panels show the reionization histories, emissivity histories, and Ly$\alpha$ forest mean transmission fractions in our \textsc{Late Start} models. As in Figure \ref{fig:Ptau_cal}, colors denote different emissivity-halo prescriptions.  The dotted curves show the original models, which are calibrated to only the mean transmission fraction, while the solid curves show the models after recalibrating to $P(<\tau_{\rm eff}^{50})$.}  
    \label{fig:Ptau_xHII}
\end{figure}

\subsubsection{The Ly$\alpha$ Forest Opacity Fluctuations}
\label{sec:taueff}

The models discussed in the previous sections were calibrated to match the observed mean transmission of the Ly$\alpha$ forest.  This calibration, however, does not guarantee agreement with the observed opacity fluctuations at $5 < z < 6$.  In fact, we find that our calibrated models in which the sources are more clustered tend to over-predict the fluctuation amplitude. We illustrate this in the top row of Figure \ref{fig:Ptau_cal} using our \textsc{Late Start} models (purple data points in Figure \ref{fig:source_z05_Ngh}). All panels show $P(< \tau^{50}_{\rm eff})$, the cumulative probability distribution function of the effective optical depth, $\tau^{50}_{\rm eff}$. Here, $\tau^{50}_{\rm eff} \equiv -\ln\langle F_{\rm Ly \alpha} \rangle_{50}$, there the average is over a pathlength of $50~h^{-1}$cMpc, and $F_{\rm Ly \alpha}$ is the continuum-normalized flux, i.e. the transmission fraction. The left, middle, and right panels correspond to $z=5.4, 5.6$, and $5.8$, respectively.  The red histograms show the observational measurements of Ref. \cite{Bosman2021}, while the other shaded regions correspond to the 10th and 90th percentiles obtained by bootstrap resampling $P(< \tau^{50}_{\rm eff})$ in the different source clustering models, as indicated in the legend.  Across redshift, the $P(< \tau^{50}_{\rm eff})$ is too wide in all of the models, indicating an overall excess in $\tau^{50}_{\rm eff}$ fluctuations for the \textsc{Late Start} scenarios.  This excess is caused by there being too many neutral islands in the $z<6$ IGM in the \textsc{Late Start} models.   

A key point here is that, for a fixed source clustering model, achieving better agreement with the observed width of $P(< \tau^{50}_{\rm eff})$ requires perturbing the reionization history to reduce the global neutral fraction at $z<6$. We illustrate this in the bottom row of Figure \ref{fig:Ptau_cal}. There, we consider the same source clustering models, but we have perturbed the \textsc{Late Start} reionization histories to achieve better agreement with the observed width of $P(< \tau^{50}_{\rm eff})$, while also maintaining good agreement with the mean transmission fraction. The new reionization histories are shown as the solid curves in the left panel of Figure \ref{fig:Ptau_xHII}, against the original histories (dashed curves).  Note that the original histories are very close to each other by design because they were used in the previous section to isolate the effects of source clustering. 

For the \textsc{$\dot{n}_{\rm ion} \propto L_{\rm UV}$} and \textsc{UV-Bright} source models, the narrowness of the observed $P(< \tau^{50}_{\rm eff})$ favors histories in which reionization begins somewhat earlier. The reason is that perturbing the end of reionization to reduce the neutral fraction at $z < 6$ results in too much overall transmission in the forest, overshooting the observed $\langle F_{\rm Ly\alpha} \rangle$ evolution. This problem  is amplified in models with larger fluctuations in the ionizing background owing to the enhanced forest transmission near the sources.  Indeed, among our three models, the \textsc{UV-Bright} source model overshoots $\langle F_{\rm Ly\alpha} \rangle$ the most when the neutral fraction is reduced at $z<6$.  We find that further perturbing the reionization history to start earlier allows us keep the neutral fraction sufficiently low to match the width of $P(< \tau^{50}_{\rm eff})$, while at the same time maintaining reasonable agreement with the $\langle F_{\rm Ly\alpha} \rangle$ evolution.  This occurs because the mean photo-ionization rate in the ionized IGM can afford to be lower at $z < 6$ if a larger fraction of reionization took place early.  These points are illustrated in Figure \ref{fig:Ptau_xHII}. Note that the fluctuation amplitude of the ionizing background is significantly lower in the \textsc{Democratic} source model. Hence, reducing the neutral fraction at $z<6$ does not result in such a sharp increase in $\langle F_{\rm Ly\alpha} \rangle$. In fact, for the illustrative example in Figure \ref{fig:Ptau_xHII}, we do not find the need to perturb the earlier phase of reionization in the \textsc{Democratic} model.

\begin{figure}
    \centering
    \includegraphics[width=0.95\linewidth]{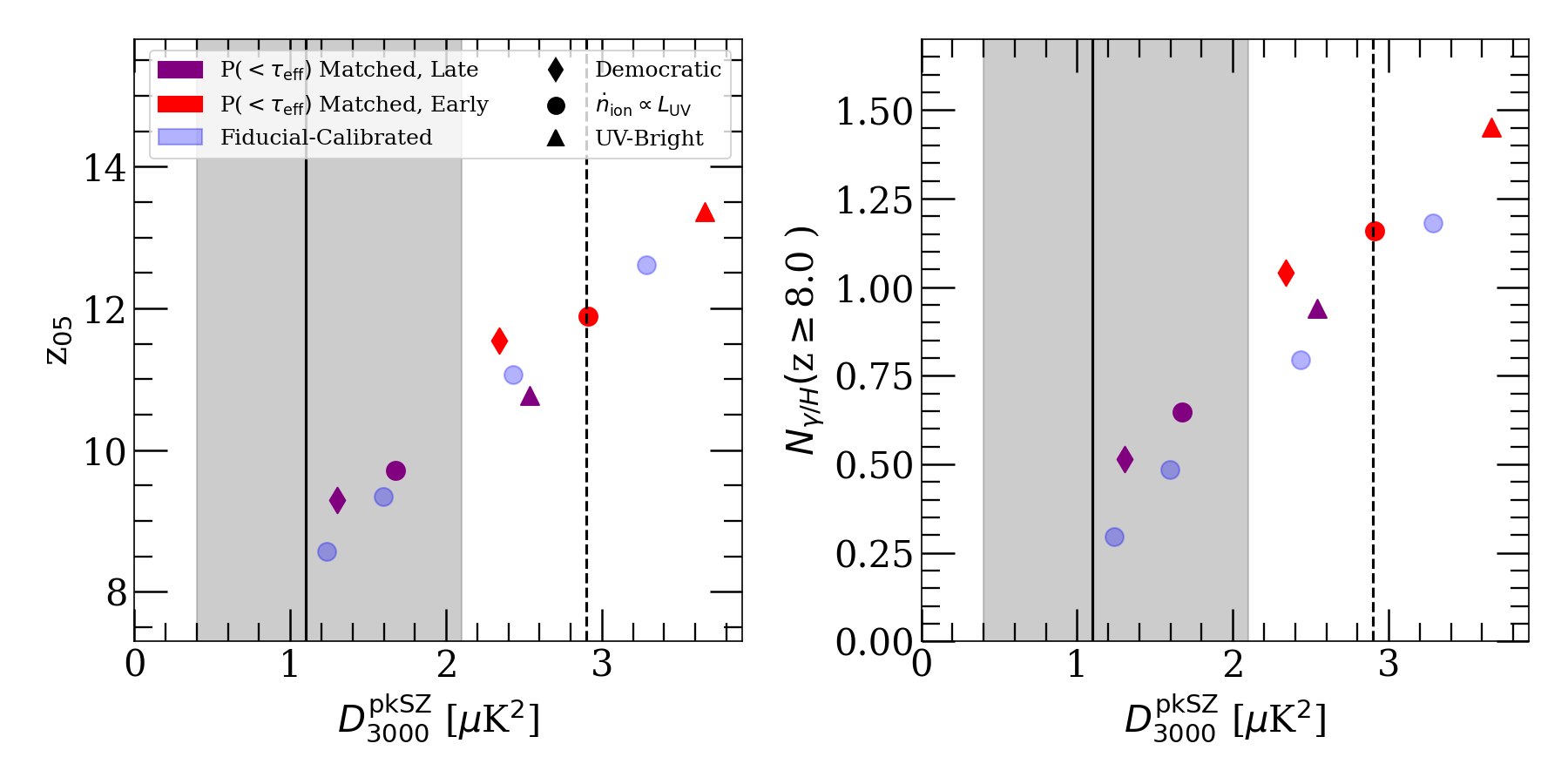}
    \caption{The same as Figure \ref{fig:source_z05_Ngh}, but for models re-calibrated to improve agreement with the Ly$\alpha$ forest opacity fluctuations.   The purple and red points correspond to the \textsc{Late} and \textsc{Early} reionization histories, while the plot markers denote different emissivity-halo prescriptions. For reference, the light blue circles show a selection of the \textsc{Fiducial-Calibrated} models from \S \ref{sec:kSZhighzlim}, illustrating the sharp correlations found in that section.  The alterations to the reionization histories driven by the recalibrations (see Fig. \ref{fig:Ptau_xHII}) tend to push models more in line with the \textsc{Fiducial-Calibrated} models. These results show that including the observed Ly$\alpha$ forest opacity fluctuations among the constraints on the end of reionization helps break the degeneracy in $D_{3000}^{\rm pkSZ}$ between the early reionization history and source clustering. }
    \label{fig:Ptau_z05}
\end{figure}

In Figure \ref{fig:Ptau_z05}, we illustrate how these re-calibrations to include the $P(<\tau_{\rm eff}^{50})$ constraints affect the relationships between $D^{\rm pkSZ}_{3000}$, $z_{05}$, and $N_{\gamma /H}^{8}$.  Here, we include a re-calibrated set of \textsc{Early} reionization models (not shown in Figure \ref{fig:Ptau_xHII}; see the green points in Figure \ref{fig:source_z05_Ngh}).  The re-calibrated \textsc{Late} and \textsc{Early} models are shown as purple and red points respectively. The different plot symbols correspond to the variation of source clustering model, as indicated in the legend. For reference, the light blue circles show a sub-sample of the \textsc{Fiducial-Calibrated} models (with $\dot{n}_{\rm ion} \propto L_{\rm UV}$) discussed in \S \ref{sec:kSZhighzlim}.  These points are reproduced here to show the trends discussed previously.   

In Figure \ref{fig:source_z05_Ngh}, we saw that changing the source clustering for fixed reionization history modulated $D^{\rm pkSZ}_{3000}$ while preserving $z_{05}$ and $N_{\gamma / H}^{8}$, spoiling the tight scalings discussed in \S \ref{sec:kSZhighzlim}.  Here, we find that including the $P(<\tau_{\rm eff}^{50})$ constraints perturbs the reionization history in a way that somewhat recovers the previously noted scalings between pkSZ power, $z_{05}$, and $N_{\gamma / H}^{8}$.  Comparing directly to Figure \ref{fig:source_z05_Ngh}, we see that the \textsc{$\dot{n}_{\rm ion} \propto L_{\rm UV}$} and \textsc{UV-Bright} models -- especially the latter -- move upward and to the right, i.e. increasing in $D^{\rm pkSZ}_{3000}$, $z_{05}$, and $N_{\gamma / H}^{8}$, for the reasons described above. This recovers somewhat the tight scaling relations found in Figure \ref{fig:z05vsD3k}. We caution that the particular models shown here are merely illustrative. They nonetheless serve to highlight how including the $\tau_{\rm eff}$ fluctuation amplitude in the emprical constraints can help break the $D^{\rm pkSZ}_{3000}$ degeneracy between reionization history and morphology.   

\subsubsection{The Shape of the pkSZ Power Spectrum}
\label{sec:pkSZshape}

Another potential way forward is to consider the pkSZ power spectra at additional scales.  Observationally, this has proved challenging due to primary CMB anisotropies increasingly dominating at $\ell \lesssim 3000$ and foregrounds dominating at $\ell \gtrsim 3000$.  However, given recent developments in foreground mitigation methods, Ref. \cite{Jain2023-tk} forecast the ability to isolate the pkSZ signal on scales $\ell \in [2500, 5000]$ in the near future.  Measurements at these additional scales could give us crucial information on the clustering of ionizing sources, the utility of which we explore in this section. 

\begin{figure}
    \centering
    \includegraphics[width=0.8\linewidth]{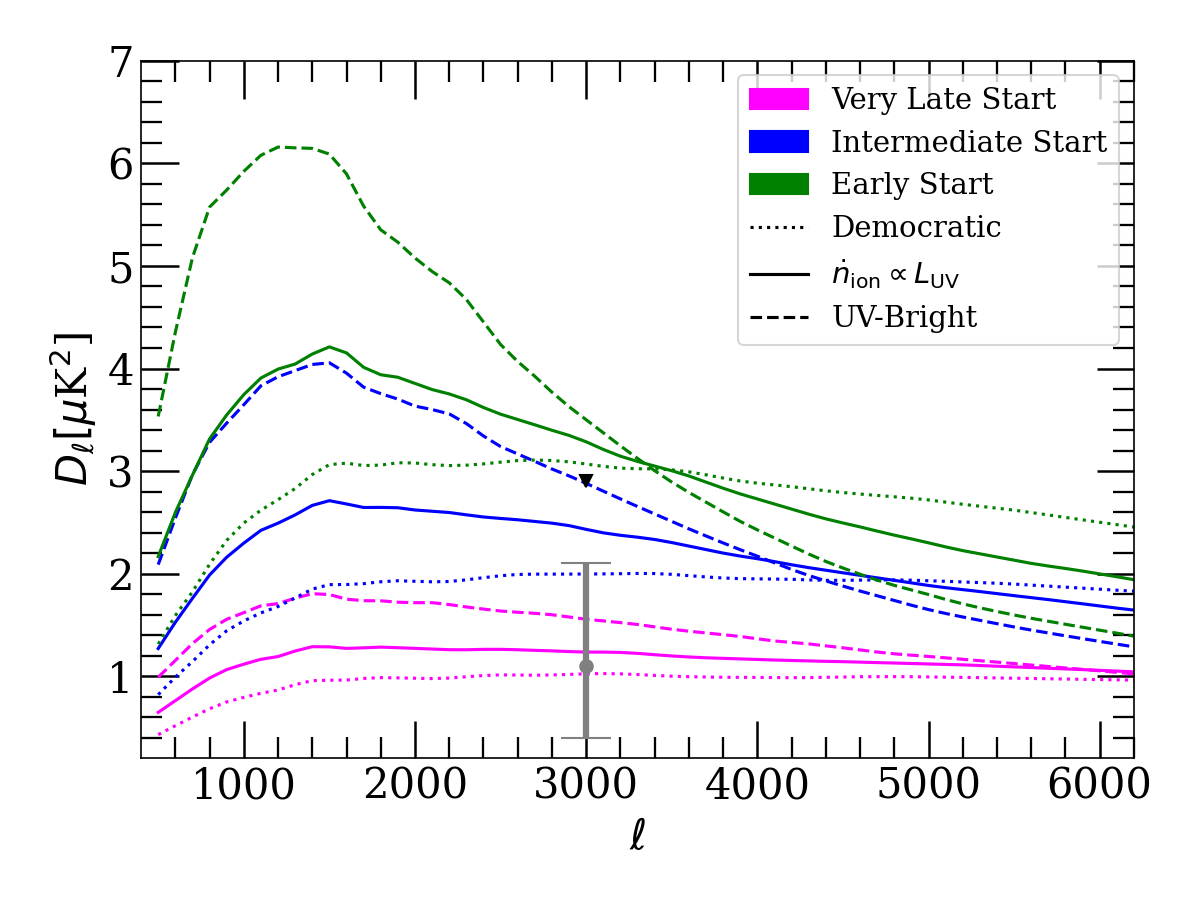}
    \caption{ The interplay between reionization history and morphology in shaping the pkSZ angular power spectrum.  The colors denote different reionization histories, while the line styles denote different emissivity-halo prescriptions.  The \textsc{Democratic} source models (least biased sources) have much flatter power spectra.  Conversely, the \textsc{UV-Bright} models (highly biased sources) have the steepest slope. These results suggest that the degeneracy between reioinzation history and source clustering can be mitigated by examining the slope and amplitude of pkSZ power spectrum.}
    \label{fig:source_xH_D3k}
\end{figure}

To start, we plot the pkSZ angular power spectrum, $D^{\rm pkSZ}_{\ell}$, in Figure \ref{fig:source_xH_D3k} for several combinations of source models and reionization histories. \textsc{Democratic}, \textsc{$\dot{n}_{\rm ion} \propto L_{\rm UV}$}, and \textsc{UV-Bright} prescriptions are represented by dotted, solid and dashed lines, respectively, with different colors denoting different reionization histories according to the legend.  It is apparent that both the slope and amplitude of $D^{\rm pkSZ}_{\ell}$ around $\ell = 3000$ are dependent on the source clustering, with the \textsc{Democratic} models having flat slopes, and the \textsc{UV-Bright} models having the steepest slopes.  These differences arise from the different morphologies of ionized regions between source models. At fixed ionized fraction, \textsc{UV-Bright} models have the largest ionized bubbles on average, so they generate more pkSZ power on larger scales (lower $\ell$), compared to the other source models.   A complication, however, is that longer reionization histories strengthen the impact of morphology on the $D^{\rm pkSZ}_{\ell}$ slope.  This is the most noticeable among the \textsc{UV-Bright} models (dashed curves).  Intuitively, extending the reionization history means that CMB photons pass through more ionized structures which imprint more temperature anisotropies, increasing the overall amplitude of $D^{\rm pkSZ}_{\ell}$.  However, for different reionization morphologies, the pkSZ power at different scales will increase by different amounts, altering the $D^{\rm pkSZ}_{\ell}$ slope in a morphology-dependent way.  This intuition suggests the that history-source degeneracy can be mitigated by examining the slope and amplitude of $D^{\rm pkSZ}_{\ell}$ simultaneously.   

\begin{figure}
    \centering
    \includegraphics[width=0.8\linewidth]{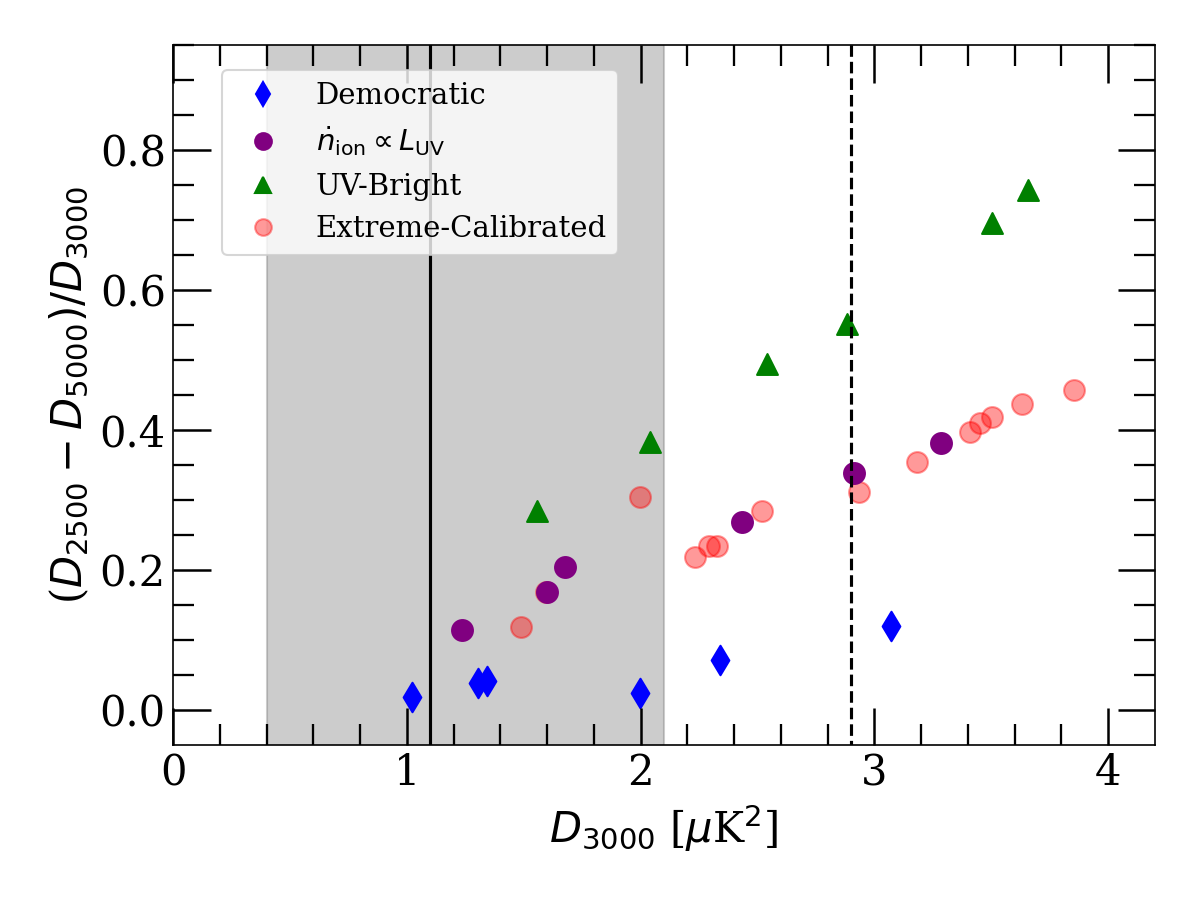}
    \caption{Relationship between the slope of the pkSZ power spectrum and its amplitude around $\ell = 3,000$ for a set of models spanning a wide range of reionization histories and source clustering prescriptions. (See main text for details.)  On the ordinate, the slope over $2500 \leq \ell \leq5000$ is normalized by the value $D_{3000}$.  Different source clustering prescriptions segregate into different trends as indicated by the marker styles. If future CMB experiments are able to measure the pkSZ power at multiple angular scales, then the slope of the power spectrum can help break the degeneracy between reionization history and source clustering, especially if $D_{3000}$ turns out to be on the high side of the values allowed by existing measurements.}
    \label{fig:slopeDl}
\end{figure}

In Figure \ref{fig:slopeDl}, we plot the normalized power spectrum slope, defined here to be $(D_{2500} - D_{5000})/D_{3000}$, as a function of $D_{3000}$ for a sample of models spanning different reionization histories and source clustering prescriptions.\footnote{The $\ell$ values are selected to align with the range specified in the aforementioned forecasts of Ref. \cite{Jain2023-tk}, but we do not expect our results to be sensitive to these choices.}  This visualizes how steep the slope is relative to the power amplitude.  \textsc{Democratic}, \textsc{ $\dot{n}_{\rm ion} \propto L_{\rm UV}$} and \textsc{UV-Bright} prescriptions are represented by blue diamonds, purple circles and green triangles, respectively. We additionally include \textsc{Extreme-Calibrated} models (with fixed \textsc{$\dot{n}_{\rm ion} \propto L_{\rm UV}$} sources) as the faded red circles.   The key result of Figure \ref{fig:slopeDl} is that different source prescriptions follow different trend lines.  For example, \textsc{Democratic} models fall quite low across the $D_{3000}$ range, reflecting their flat $D_{\ell}$ slopes.  In contrast, \textsc{$\dot{n}_{\rm ion} \propto L_{\rm UV}$} and \textsc{UV-Bright} models show distinct trends of steepening slopes with increasing $D_{3000}$, with the latter's power spectra being around twice as steep as the former's for the same $D_{3000}$.   Reassuringly, these trends seems to hold across the range of histories considered, including both \textsc{P($<\tau_{\rm eff}^{50}$) Matched} models and even the majority of \textsc{Extreme-Calibrated} models.\footnote{The notable exception in the \textsc{Extreme-Calibrated} set at ($\sim 2\mu$K$^2$, $\sim$ 0.3) is a history where reionization starts with an extremely rapid growth in emissivity.  In such a scenario, small ionized structures at very low $x_{\rm HII}^{\rm V}$ do not persist long and imprint less power at small scales, resuting in a steeper slope.}  If indeed future experiments are able to measure the pkSZ signal over $\ell\in [2500, 5000]$, then the slope of the power spectrum may help break the degeneracy between reionization history and source clustering. We note, however, that the usefulness of the slope is limited by the overall power amplitude; the differences in the slope increase with $D_{3000}$.  Indeed, our results suggest that the slope may be of limited utility if $D_{3000}$ turns out to be low, e.g. near the central value of the existing measurements by Refs. \cite{Reichardt2020, Beringue2025-ka}. 

\section{Discussion}
\label{sec:discussion}

All of the models in our main results adopt a temporally fixed emissivity-halo relationship.  For instance, in the $\dot{n}_{\rm ion} \propto L_{\rm UV}$ models, the emissivity assigned to each source is assumed proportional to its UV luminosity which, in turn, is mapped uniquely to a halo mass through abundance matching. In this sense, although the overall normalization of the emissivity evolves with time (owing to our calibration to the forest data), the emissivity-halo connection remains fixed.  It is easy to imagine, however, that the emissivity-halo connection evolves during reionization.  For example, smaller/fainter sources might start reionization before giving way to more massive galaxies as the former succumb to feedback processes \cite{Park2013-bc, 2007MNRAS.376..534I, 2021MNRAS.507.6108O}. There is also the issue of stochastic star formation, especially at the highest redshifts of interest. Recent observations provide evidence for extreme burstiness at these epochs \cite{Endsley2023, Kravtsov2024-la, Looser2023-kh, Kokorev2025-qa, Stark2025-rh, Cole2023-we, Topping2024-mo, Ciesla2023-ut}.  In \S's \ref{sec:evolvingsources} and \ref{sec:stochastic}, we use illustrative models to argue that such effects are unlikely to change our broad conclusions.  In \S \ref{sec:comparison}, we compare the results discussed above to the findings of previous studies.  

\subsection{Models with evolving emissivity-halo connection}
\label{sec:evolvingsources}
\begin{figure}
    \centering
    \includegraphics[width=0.8\linewidth]{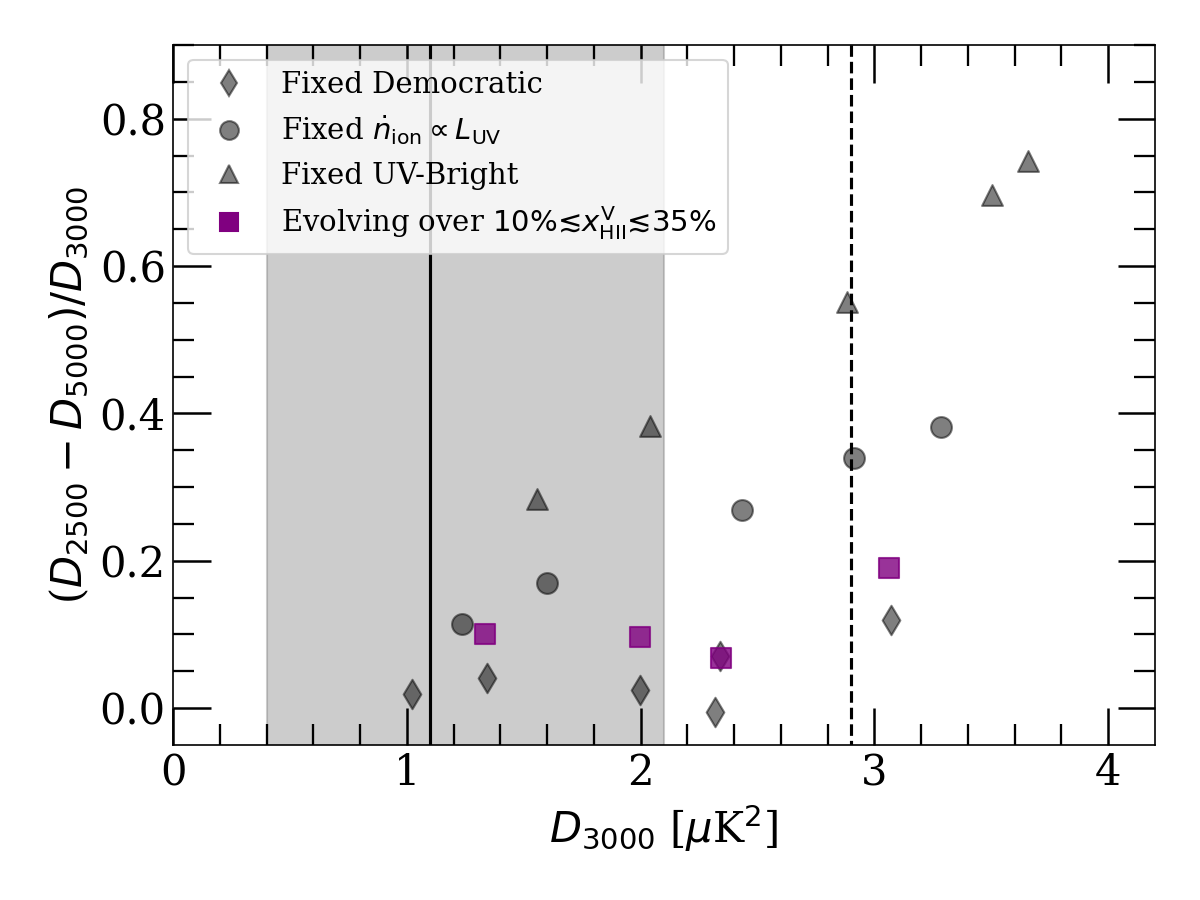}
    \caption{The same as Figure \ref{fig:slopeDl}, but for models in which the emissivity-halo connection evolves in time.  For reference, the gray data points reproduce some of the same models shown in Figure \ref{fig:slopeDl}.  The purple squares show new models where reionization begins with the \textsc{Democratic} source prescription, but evolves into the \textsc{UV-Bright} prescription during the interval $10\% \lesssim x_{\rm HII}^{\rm V} \lesssim 35\%$.  This time scale is extremely rapid and chosen to maximize the change in $D_{\ell}$ for illustrative purposes.   Even in this extreme case, we see that the evolving source models are only modestly different than the \textsc{Democratic} models, with somewhat steeper slopes and similar amplitude. The takeaway is the slope of the pkSZ power around $\ell = 3, 000$ appears to be most sensitive to the clustering of the sources that drive the first $\sim 30\%$ of reionization. Since the evolving models start out identical to the \textsc{Democratic} models, they are most similar in this plot.}   
    \label{fig:evolving_sources}
\end{figure}

Consider a scenario in which reionization is started by smaller halos, but finished by more massive halos. We develop a simple ansatz to obtain a rough upper limit for how much this kind of evolving emissivity-halo connection would change our results.  In the ansatz, the sources start out identical to our \textsc{Democratic} model and evolve into the \textsc{UV-Bright} model.  To achieve this, we adopt the mapping $\dot{n}_{\rm ion} \propto L_{\rm UV}^{\beta}$ and assume that $\beta$ evolves from 0 to 1 linearly in redshift.  Additionally, the cutoff in UV-magnitude above which sources have zero emissivity, $M_{\rm UV}^{\rm Cut}$, evolves from -10 to -18 linearly over the same time interval.  (Note that $\beta = 0, M_{\rm UV}^{\rm Cut} = -10$ and $\beta = 1, M_{\rm UV}^{\rm Cut} = -18$ correspond to our \textsc{Democratic} and \textsc{UV-Bright} models, respectively.)  As shown in Figure \ref{fig:Dl_vs_xandz}, $D_{\ell}$ receives most of its contribution at $x_{\rm HII} \lesssim 40\%$. Hence, we chose to vary these parameters quickly over $x_{\rm HII} \sim 10\% - 35\% $, i.e. the transition from \textsc{Democratic} to \textsc{UV-Bright} occurs before $x_{\rm HII} \sim 35\%$.  This rapid evolution corresponds to timescales of $\sim$ 170 - 240 Myr, for our shortest and longest \textsc{Fiducial-Calibrated} histories, respectively. We emphasize that this extreme model is for illustrative purposes, and we use it here to gauge a rough upper limit for the effect. 

Figure \ref{fig:evolving_sources} considers the slope of the pkSZ power spectrum versus $D_{\ell}$ (as in Fig. \ref{fig:slopeDl}), this time showing results from the models with evolving emissivity-halo connection as the purple squares.  For comparison, the gray diamonds, circles and squares represent a selection of the \textsc{Democratic}, \textsc{ $\dot{n}_{\rm ion} \propto L_{\rm UV}$}, and \textsc{UV-Bright} models from Figure \ref{fig:slopeDl}.  We see that the evolving models exhibit modestly steeper slopes than the \textsc{Democratic} models, while preserving similar $D_{3000}$.  Notably, despite the extremely quick evolution toward \textsc{UV-Bright} (for the last $\sim$ 60\% of reionization, the sources are nearly identical to those in the \textsc{UV-Bright} models), the evolving models present mostly like the \textsc{Democratic} ones. A more realistic/slower evolution in the emissivity-halo connection would yield slopes even closer to the \textsc{Democratic} results.  This finding supports the idea, raised in \S \ref{sec:zstart} and \S \ref{sec:clustering}, that the pkSZ power spectrum probes the source properties during the first $\sim 30 \%$ of reionization.  The key takeaway from this section is that the slope of the pkSZ power around $\ell = 3,000$ appears to be most sensitive to the clustering of the sources that drive the first $\sim 30\%$ of reionization, consistent with \S \ref{sec:clustering}. Any changes to the emissivity-halo connection that occur later have only a modest effect on the shape of the power spectrum.

\subsection{Stochastic Star Formation}
\label{sec:stochastic}

Stochastic star formation has been invoked to explain the abundance of UV-bright galaxies at $z>10$ \cite{2023MNRAS.518.6011D, 2023MNRAS.523.1009B, 2023ApJ...948L..14C, 2023ApJ...951L...1P, 2023ApJ...954L..46L, 2024MNRAS.527.5004M, Finkelstein2024, Adams2024, 2025arXiv250100984W, Furlanetto2021-wr, Mirocha2022-mw, Sun2023-dj, Sun2023-ju,Sun2024-cb, Andalman2024-lm, Gelli2024-ii}.  Indeed, the spectral energy distributions of UV-bright galaxies at this epoch show evidence of ongoing starbursts \cite{Endsley2023, Kravtsov2024-la, Looser2023-kh, Kokorev2025-qa, Stark2025-rh, Cole2023-we, Topping2024-mo, Ciesla2023-ut}.   While we already use the UV luminosity functions derived from JWST observations, stochastic star formation histories can complicate the emissivity-halo connection in a way not captured by our abundance matching model.  What if a significant fraction of the UV-bright galaxies in the observed UV luminosity function actually reside within low-mass halos that happen to be undergoing intense bursts of star formation, altering the effective clustering strength of the sources? In this section, we use a simple model to test such a scenario. 

The basic features of our simplistic approach are motivated by the semi-analytic model of Ref. \cite{Furlanetto2021-wr}. Their model adopts a timescale of $\sim 5-30$ Myr between star formation and  stellar feedback episodes.  Star formation proceeds uninhibited over this time scale, resulting in a massive starburst.  In smaller halos, the associated feedback is strong enough to drive gas out of the galaxy and quench star formation.  For these halos, the model of Ref. \cite{Furlanetto2021-wr} predicts a change in the rest-frame UV magnitude, $M_{\rm UV}$, of $\pm$ 1 over starburst-quench cycles, with bursty galaxies having higher than average SFR's $\sim 25\%$ of the time.  On the other hand, more massive halos, with their deeper gravitational wells, are able to hold onto their gas reservoirs after a feedback event, so they instead undergo steady state star formation.  The halo mass marking the transition between these two regimes depends on the assumed small-scale efficiency of star formation and the strength of feedback. Ref. \cite{Furlanetto2021-wr} estimate the maximum halo mass that can host a starburst galaxy (SBG) to be $M_{\rm max}^{\rm SBG} \sim 10^{10}-10^{10.5}$ M$_{\odot} h^{-1}$.  Here, we would like to quantify (roughly) the maximum effect that we expect stochastic star formation to have on our results.  We therefore adopt the high end of this range, $M_{\rm max}^{\rm SBG} = 10^{11}$ M$_{\odot} h^{-1}$.\footnote{For reference, in our abundance matching scheme from the previous sections, halos of this mass host galaxies with M$_{\rm UV} \sim$ -20 and -21.5 at $z=6$ and 10, respectively.}  Above $z=11$ we have no halos this massive.  In other words, with our assumed $M_{\rm max}^{\rm SBG}$, all of our $z>11$ sources are candidates for SBGs. 

In our simplistic scheme, we begin with the \textsc{UV-Bright} model described in \S \ref{sec:clustering} and the same halo mass-UV magnitude relation from the abundance matching procedure of the previous sections. At each snapshot of our hydrodynamics simulation (corresponding to $\Delta t \approx 10$ Myr), a random fraction, $f_{\rm SBG}$, of all halos with masses $\leq  M_{\rm max}^{\rm SBG}$ are selected to be SBGs.  We randomly swap the UV magnitude of each of these SBGs with a non-SBG that is $\Delta$M$_{\rm UV}$ $\sim1$ brighter.  This process is repeated at every hydrodynamical snapshot with a new random selection of SBGs and swappings.  We note that the timescale of $\sim$10 Myrs on which our SBGs flicker is consistent with the expected duration of starburst events in Ref. \cite{Furlanetto2021-wr} and other works.  Crucially, our approach preserves the {\it measured} UV luminosity function while implementing starburst events -- albeit in a crude way.  In what follows, we examine two stochastic star formation models.  The first assumes $f_{\rm SBG}=25\%$, in line with Ref. \cite{Furlanetto2021-wr},  and is termed \textsc{Stochastic-SF: $f_{\rm SBG}=25\%$}.  The second takes a more extreme value of $f_{\rm SBG}=50\%$ and is labeled \textsc{Stochastic-SF: $f_{\rm SBG}=50\%$}.  We note $f_{\rm SBG}=50\%$ is the maximum value allowed in this approach for $z\gtrsim11$.  Because of our choice of $M_{\rm max}^{\rm SBG}$, all galaxies in this epoch are SBG candidates, so $f_{\rm SBG}>50\%$ means there are more SBGs than non-SBG's, which would ruin our swapping scheme.  Since the global ionizing emissivities of both \textsc{Stochastic-SF} and \textsc{UV-Bright} models are identical by construction, any differences in their pkSZ power owe solely to the changes in effective source clustering introduced by stochastic star formation, isolating exactly the effect we aim to test. 

\begin{figure}
    \centering
    \includegraphics[width=0.95\linewidth]{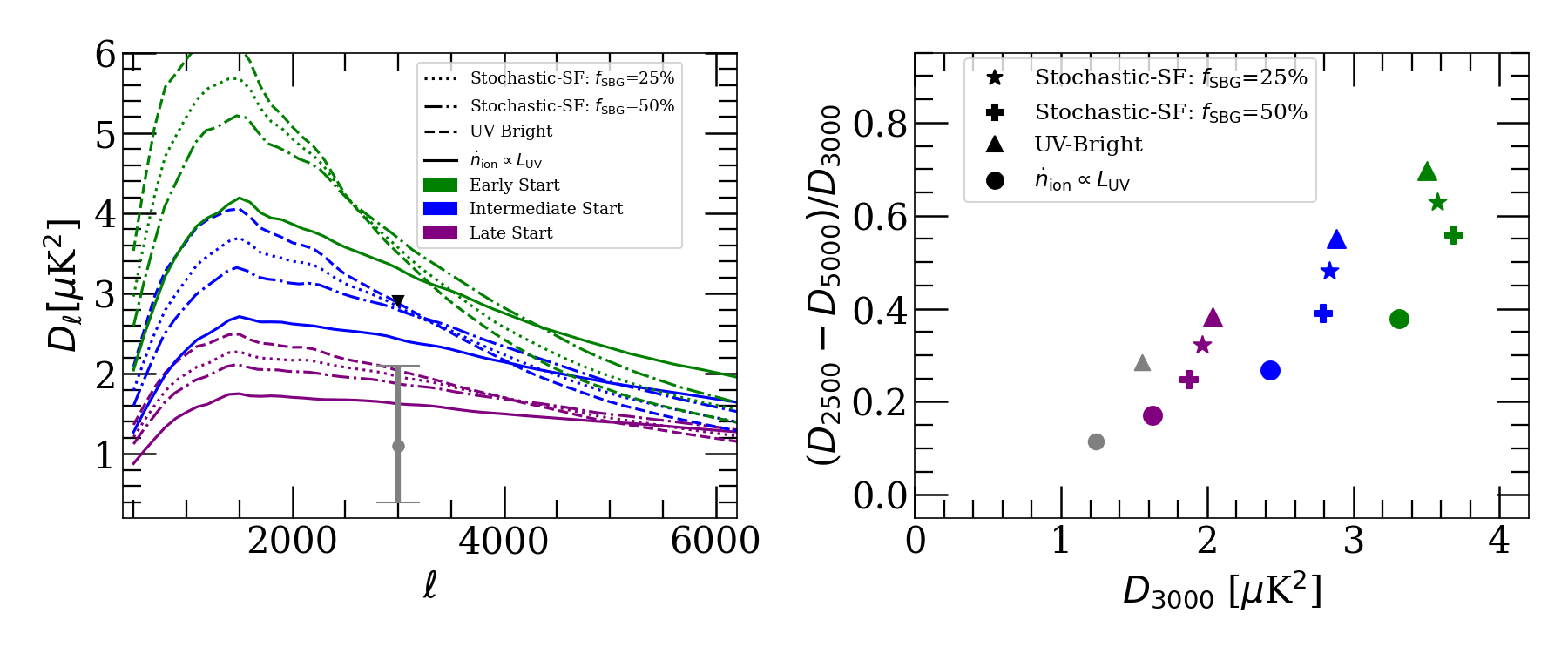}
    \caption{Testing the impact of stochastic star formation in < 10$^{11}$ M$_{\odot} h^{-1}$ halos on the pkSZ power spectra.  \textit{Left:} $D_{\ell}$ for our \textsc{Stochastic-SF: $f_{\rm SBG}=25\%$} (dotted) and \textsc{Stochastic-SF: $f_{\rm SBG}=50\%$} models (dot-dashed) compared against \textsc{UV-Bright} (dashed) and \textsc{$\dot{n}_{\rm ion} \propto L_{\rm UV}$} (solid) source models for various reionization histories denoted by color.  Both stochastic star formation models can be viewed as variations of the \textsc{UV-Bright} model, but with stochasticity added using a prescription motivated by the semi-analyic model of Ref. \cite{Furlanetto2021-wr} (see main text for details). \textit{Right:} the normalized slope of the pkSZ power as a function of power amplitude at $\ell = 3,000$. The stars and crosses represent the \textsc{Stochastic-SF: $f_{\rm SBG}=25\%$} and \textsc{Stochastic-SF: $f_{\rm SBG}=50\%$} models respectively. The similarity between the \textsc{Stochastic-SF: $f_{\rm SBG}=25\%$} and \textsc{UV-Bright} results suggest that stochasticity in a moderate fraction of galaxies likely has only a modest effect on the broad conclusions of the previous sections.  However, the similarity between the \textsc{Stochastic-SF: $f_{\rm SBG}=50\%$} and \textsc{$\dot{n}_{\rm ion} \propto L_{\rm UV}$} trend lines shows that a strong prevalence of stochastic star formation is partially degenerate with the effects of weaker source clustering. }
    \label{fig:SBG}
\end{figure}

Figure \ref{fig:SBG} shows how stochastic star formation affects the pkSZ power spectra.  In the left panel we compare $D_{\ell}$ in our \textsc{Stochastic-SF: $f_{\rm SBG}=25\%$} (dotted) and \textsc{Stochastic-SF: $f_{\rm SBG}=50\%$} models (dot-dashed) against the corresponding \textsc{UV-Bright} models (dashed).  For comparison, we also show a set of \textsc{$\dot{n}_{\rm ion} \propto L_{\rm UV}$} models as the solid curves.  Different colors denote different global reionization histories, as indicated in the legend.  We can see the power spectra for each \textsc{Stochastic-SF: $f_{\rm SBG}=25\%$} model is only slightly flatter than the corresponding \textsc{UV-Bright} model.  In fact, around $\ell = 3,000$, these two sets of models are quite similar. This is emphasized in the right panel of Fig. \ref{fig:SBG}, where we plot the normalized slope versus the amplitude of power at $\ell = 3,000$  (the same quantities plotted in Fig. \ref{fig:slopeDl}).  Our \textsc{Stochastic-SF: $f_{\rm SBG}=25\%$} models (stars) have only $\sim$ few \% different $D_{3000}$, and $\sim 10-15\%$ lower values of ($D_{2500}-D_{5000}) / D_{3000}$, compared to their corresponding \textsc{UV-Bright} models (triangles).  Overall, when implementing stochasticity into a moderate fraction of galaxies in the \textsc{UV-Bright} models, we get results quite similar to the original \textsc{UV-Bright} models.  However, the $D_{\ell}$ slope continues to flatten as $f_{\rm SBG}$ increases.  In fact, for histories with a \textsc{Late Start} or \textsc{Intermediate Start}, the \textsc{Stochastic-SF: $f_{\rm SBG}=50\%$} models (pluses) have only slightly higher values of ($D_{2500}-D_{5000}) / D_{3000}$ than the \textsc{$\dot{n}_{\rm ion} \propto L_{\rm UV}$} models (circles), despite the latter allowing galaxies fainter than $M_{\rm UV}=-18$ to contribute to reionization.  This suggests an extreme prevalence of stochastic star formation can be partially degenerate with the inclusion of fainter galaxies in the effective clustering of ionizing sources and thus, the $D_{\ell}$ slope.  A more complete exploration of this degeneracy between source modeling and stochasticity is left to future work. 

The main takeaway from this test is that stochastic star formation among a moderate fraction of early galaxies probably has only a modest effect on the conclusions of the previous section.  However, a very high stochastic fraction flattens the slope of $D_{\ell}$ significantly. Interpreting future measurements the pkSZ power spectrum slope may require modeling the effects of stochastic star formation.

\subsection{Comparison to Previous Results}
\label{sec:comparison}

That the majority of $D_{3000}^{\rm pkSZ}$ is sourced at $x_{\rm HII}^{\rm V} \lesssim 40\%$ was also observed by Refs. \cite{Mesinger2012-ad, Park2013-bc, Gorce2020-gx}.  Using a smaller set of RT simulations, Ref.~\cite{Park2013-bc} found that the fraction of power contributed at $x_{\rm HII}^{\rm V} \lesssim 40\%$ decreases if halos with masses $\sim 10^9 M_{\odot}$ drive reionization.  However, they also found that this fraction remains above $50\%$, and does not decrease further, even when including contributions from atomic cooling halos down to masses of $10^8 M_{\odot}$.  Ref.~\cite{Mesinger2012-ad} found a similar result using the semi-numeric code \textsc{21cm-fast}.  In that work, the exceptions were models in which the ionizing photon mean free path {\it in ionized gas} remains short enough to prevent the growth of ionized bubbles much larger than $\sim 10$Mpc (i.e. until they merge with neighboring bubbles).  In this scenario, smaller bubbles persist until much higher $x_{\rm HII}^{\rm V}$ and continue to contribute significantly to the pkSZ power at $\ell = 3000$ throughout reionization.  Despite the differences in modeling between these works, they generally agree that in standard reionization scenarios driven by galaxies in atomic cooling halos, a majority of $D_{3000}^{\rm pkSZ}$ is sourced at $x_{\rm HII}^{\rm V} \lesssim 40\%$. 

In contrast with our findings in \S \ref{sec:zstart}, Ref. \cite{Park2013-bc} found little change in $D_{3000}^{\rm pkSZ}$ with earlier starts to reionization.  We argue that this discrepancy with our results is likely due to the relationship between their reionization histories and source models.  For example, their ``L2M1J1'' model begins at $z\approx25$ with the first $\approx20\%$ of reionization driven by ionizing sources within minihalos down to masses $\sim10^5 M_{\odot}$.  As discussed in \S \ref{sec:clustering}, less clustered sources generate fewer $\ell\sim 3000$ scale ionized bubbles at low-$x_{\rm HII}^{\rm V}$, reducing contributions to $D_{3000}^{\rm pkSZ}$ from those epochs.  Indeed, the red curve in Figure 8 of Ref. \cite{Park2013-bc} shows quite weak contributions to $D_{3000}^{\rm pkSZ}$ for this minihalo driven scenario all the way down to $z\sim11$.  At $z < 10$, when atomic cooling halos take over as the primary drivers of reionization, the higher $x_{\rm HII}^{\rm V}$ means their ionized regions grow larger than $\ell\sim3000$ angular scales somewhat sooner than they otherwise would.  This results in the model with minihalos having slightly lower $D_{3000}^{\rm pkSZ}$ contributions from $z\lesssim10$, as seen in their Figure 8.   It is possible that the slight loss of power from $z\lesssim10$ offsets the mild early gains and makes $D_{3000}^{\rm pkSZ}$ effectively insensitive to the start of reionization.  We emphasize this seems to be a result of the less clustered sources driving the earlier starts to reionization, not of minihalos in particular. Ref \cite{Park2013-bc} sees similar behavior when comparing a scenario driven solely by high-mass atomic cooling halos $(\geq 2.2 \times 10^{9} M_{\odot})$ against a model that includes sources within low-mass atomic cooling halos $(\geq10^8 M_{\odot})$.    This scenario serves as an important counterexample to our finding that $D_{3000}^{\rm pkSZ}$ is sensitive to the start of reionization.

Some recent analyses of the CMB temperature and lensing data within the \textsc{$\Lambda$CDM} model find a preference for a larger CMB optical depth of $\tau_{\rm es}\sim0.09$ when the Baryon Acoustic Oscillation data from DESI \cite{DESI-Collaboration2025-yg} is included \cite{Jhaveri2025-wt, Sailer2025-ie}\footnote{Such analyses do not include the low-$\ell$ EE polarization data from \textit{Planck}, which is the primary observational evidence for $\tau_{\rm es}\sim0.06$}. However, using a suite of semi-numeric simulations with the end of reionization set to be consistent with Ly$\alpha$ forest constraints, Ref. \cite{2025arXiv250515899C} pointed out such high $\tau_{\rm es}$ is in $\gtrsim2 \sigma$ tension with the low value of $D_{3000}^{\rm pkSZ}$ measured by SPT.  We emphasize that this tension only occurs because of the strict constraints on the end of reionization from the Ly$\alpha$ forest.  The scenario described in the previous paragraph -- where reionization begins early, driven by weakly clustered sources -- could plausibly satisfy all three constraints.  The early start to reionization yields a high $\tau_{\rm es}$, and it is possible that the rarity of  $\ell \sim 3000$ scale ionized regions generated by the minihalo sources keeps $D_{3000}^{\rm pkSZ}$ low, even when reionization ends around $z\sim5.5$ (as required by the Ly$\alpha$ forest).  This discussion further underscores the need to clarify the role of weakly clustered sources at the beginning of reionization to properly interpret $D_{3000}^{\rm pkSZ}$ (see Ref. \cite{Park2013-bc}). As discussed in \S \ref{sec:pkSZshape}, a signature of weak source clustering is more pkSZ power at larger $\ell$, so measurements from future small scale CMB experiments are a promising way to gain insights into the earliest sources of reionization.

\section{Conclusion}
\label{sec:conclusion}
Over the past decade, advances in Ly$\alpha$ forest observations at $z > 5$ have brought reionization's last phases into focus.  Indeed, forest measurements now provide more than a simple lower limit on the end of reionization; they place stringent constraints on its timing (see e.g. the discussion in \cite{2025arXiv250515899C}). These constraints will continue to improve as more high-$z$ quasars are discovered in wide-field cosmological surveys and followed up to obtain high signal-to-noise spectra. On the other hand, future CMB experiments have the potential to provide robust constraints on the highly uncertain earlier phases of reionization.  Better constraining the start of reionization would, in turn, provide valuable insights into the nature and ionizing properties of JWST sources detected at these epochs. We have used a suite of radiative transfer simulations to explore how the pkSZ power spectrum could be used as such a window when combined with Ly$\alpha$ forest constraints on reionization's end. Our main results are summarized as follows: 

\begin{itemize}
    \item For the galaxy-sourced models considered in this paper (with $M_{\rm min} \geq 10^9h^{-1}$M$_\odot$), The pkSZ power at $\ell \sim 3,000$ receives most of its contribution from the first $40\%$ of reionization.  
    
    \item For a fixed model of the ionizing source clustering, when the timing of reionization's end is constrained by measurements of the Ly$\alpha$ forest mean transmission at $z=5-6$, the amplitude of the pkSZ power spectra, $D^{\rm pkSZ}_{3000}$, strongly correlates with the beginning epoch of reionization (defined here to be the redshift at which $x_{\rm HII} = 5$\%, $z_{05}$).  Imposing physically motivated limits on how quickly the global ionizing emissivity can grow, existing and future pkSZ measurements can be translated to model-dependent upper limits on the cumulative number of ionizing photons emitted per hydrogen atom by the high-$z$ source population, $N_{\gamma/H}(z\gtrsim8)$.
    
    \item Varying the assumed relationship between dark matter halos and the ionizing outputs of their host galaxies (i.e. varying the assumed source clustering) introduces a significant amount of scatter into the correlations between $D^{\rm pkSZ}_{3000}$ and both $z_{05}$ and $N_{\gamma/H}$.  This owes to a degeneracy in $D^{\rm pkSZ}_{3000}$ between the morphology of reionization (set by the clustering of ionizing sources) and the early reionization history.  Much like variations in the reionization history, various source clustering models imprint differences in the pkSZ power within the first 30\% of reionization.
    
\item
We demonstrated that the aforementioned degeneracy can be partly broken in at least two ways: (1) by incorporating additional constraints in the form of the observed spatial fluctuations in the Ly$\alpha$ forest opacity at $5 \lesssim z \lesssim 6$; (2) by incorporating information about the shape of the pkSZ power spectrum around $\ell = 3,000$. The shape of the power spectrum is sensitive to the clustering of the ionizing sources, with less clustered sources yielding flatter power spectra. This latter method is most useful in scenarios with larger amplitudes of $D^{\rm pkSZ}_{3,000}$.  
    
    \item We performed tests to explore how sensitive these conclusions are to possible time-evolution in the emissivity-halo connection, as well as highly stochastic ionizing emissions at high redshift. These tests suggest that such effects are unlikely to change our broad conclusions.  However, stochastic star formation among high fraction ($>50\%$) of galaxies can flatten the pkSZ power spectrum, complicating its interpretation.
    
\end{itemize}

Future CMB experiments such as Simons Observatory and CMB-S4, in combination with ever-improving constraints from the Ly$\alpha$ forest, have the potential to constrain the ionization state of the early IGM.  Our results suggest that a synthesis of CMB, Ly$\alpha$ forest, and JWST observations can yield deeper insights into the ionizing sources near cosmic dawn.

\acknowledgments
The authors thank Hy Trac for generously providing access to his cosmological hydrodynamics code. The authors also thank Paul Shapiro and Matt McQuinn for helpful discussions. A.D. acknowledges support from grants NASA 19-ATP19-0191 and NSF AST-2045600. All computations were made possible by NSF ACCESS allocations TG-PHY210041 and TG-PHY230063.

\appendix

\section{ Comparing pkSZ for Full and Reduced Speed of Light Runs }
\label{append:rsla_vs_fsl}

\begin{figure}
    \centering
    \includegraphics[width=0.8\linewidth]{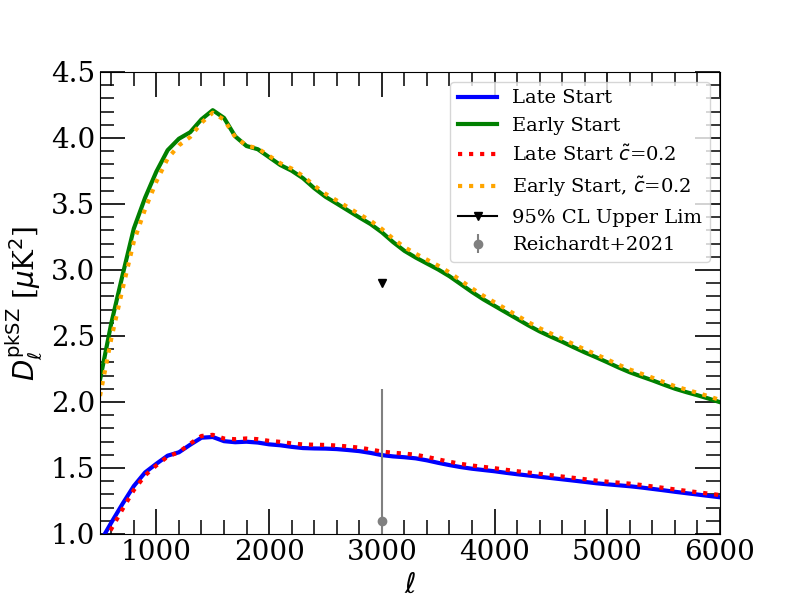}
    \caption{ Comparing the pkSZ power spectrum from full speed of light $(\Tilde{c}/c=1)$ simulations against $\Tilde{c}/c=0.2$ reduced speed of light simulations with remapped emissivities given by inversions of Eq. \ref{eq:RSLA} and Eq. \ref{eq:RSLA2}.  The pkSZ power spectra for $\Tilde{c}/c=1$ ($\Tilde{c}/c=0.2$) simulations of the early starting reionization history is given by the solid-green (orange-dotted) curve, and $D_{\ell}^{\rm pkSZ}$ for the late starting history is given by the  solid-blue (red-dotted) curve.  For both histories we find $\lesssim 1.5\%$ difference in $D_{3000}^{\rm pkSZ}$ due to our use of reduced speed of light simulations.}
    \label{fig:Dl_c02_test}
\end{figure}

To confirm the robustness of our method of remapping the emissivities from simulations employing the reduced speed of light approximation, we compare to results from simulations that use the full speed of light. Specifically, we take two reionization histories from Ref \cite{2024arXiv240902989C}, one with a late start and a second with an early start ($z_{05}\sim9.3$ and $\sim 12.5$ respectively). Both use the same \textsc{$\dot{n}_{\rm ion}\propto L_{\rm UV}$} source model used in this work and were calibrated in full speed of light simulations to agree with the measurements of the Ly$\alpha$ forest mean flux from Ref. \cite{Bosman2021}.  We find the equivalent emissivities using the inversions of Eqs. \ref{eq:RSLA} and \ref{eq:RSLA2} to use in $\Tilde{c}/c$=0.2 simulations.  In Fig. \ref{fig:Dl_c02_test} we compare the pkSZ power spectra calculated from the original full speed of light simulations against those from new $\Tilde{c}/c=0.2$ simulations with remapped emissivities.  In both the early and late starting cases the $\Tilde{c}/c=0.2$ runs gained $\sim$ 0.026 $\mu$K$^2$ in $D_{3000}^{\rm pkSZ}$ which corresponds to a difference of $\lesssim 1.5 \%$.  As such, we expect our results to be near identical  had we used full speed of light simulations.  

\bibliographystyle{JHEP}
\bibliography{reference.bib}
\end{document}